\documentclass[a4paper,11pt]{article}
\usepackage{jheppub} % for details on the use of the package, please
\usepackage[T1]{fontenc} % if needed

\usepackage{dsfont}

\newcommand{\mathbbm}[1]{\mathds{1}}

\renewcommand\H{\operatorname{H}}

\newcommand{\NNNLO}{N${}^3$LO}

\newcommand{\prog}[1]{\texttt{#1}}

\newcommand{\IncGraph}[1]{%
  \begin{minipage}[b]{\GraphSize\textwidth}%
    \centering
    \includegraphics[width=\ImageSize\textwidth,clip]{#1.eps} \\
    \ParseGraph#1:%
  \end{minipage}%
}
\newcommand{\GraphSize}{.3}
\newcommand{\ImageSize}{.8}
\def\ParseGraph#1:{#1}

\title{\boldmath Adequate bases of phase space master integrals for $gg\to h$ at NNLO and beyond}

\author{Maik H\"oschele,}
\author{Jens Hoff}
\author{and Takahiro Ueda}

\affiliation{
Institut f\"ur Theoretische Teilchenphysik,
Karlsruhe Institute of Technology (KIT) \\
76128 Karlsruhe, Germany}

\emailAdd{maik.hoeschele@kit.edu}
\emailAdd{jens.hoff@kit.edu}
\emailAdd{takahiro.ueda@kit.edu}

\abstract{
  We study master integrals needed to compute the Higgs boson production cross
  section via gluon fusion in the infinite top quark mass limit, using
  a canonical form of differential equations
  for master integrals, recently identified by Henn, which makes their solution
  possible in a straightforward algebraic way.
  We apply the known criteria to derive such a suitable basis for all the phase
  space master integrals in afore mentioned process at next-to-next-to-leading order in
  QCD and
  demonstrate that the method is applicable
  to next-to-next-to-next-to-leading order as well by solving a non-planar topology.
  Furthermore, we discuss in great detail how to find an adequate basis
  using practical examples.
  Special emphasis is devoted to master integrals which are coupled by their
  differential equations.
}

\preprint{SFB/CPP-14-38, TTP-14-20, LPN14-087}

\begin{document}
\maketitle
\flushbottom

\section{Introduction}
\label{sec:intro}

In order to study the compatibility of the assumed Higgs particle discovered
by ATLAS and CMS~\cite{Aad:2012tfa,Chatrchyan:2012ufa}
with the standard model precise theoretical predictions are required.
One of the basic physical observables is the total inclusive Higgs production cross
section which is, as is well known, dominated by gluon fusion
at the LHC.
For a long time the state of the art in fixed-order perturbative calculations
of the total inclusive Higgs production cross section
in the gluon fusion channel has been
next-to-leading-order (NLO) for electroweak corrections
and next-to-next-to-leading order (NNLO) for QCD corrections
(see ref.~\cite{Dittmaier:2011ti} for comprehensive reviews).
The latter ones have firstly been calculated in the infinite top mass limit%
~\cite{Harlander:2002wh,Anastasiou:2002yz,Ravindran:2003um}
while finite mass corrections were included in refs.%
~\cite{Marzani:2008az,Harlander:2009bw,Pak:2009bx,Harlander:2009mq,Pak:2009dg,Harlander:2009my}.

In recent years, various next-to-next-to-next-to-leading order (\NNNLO{}) QCD
approximations have become available%
~\cite{Moch:2005ky,Bonvini:2014jma}
but the full calculation remains a challenging frontier.
Some partial results have been obtained with full dependence on
the partonic center-of-mass energy
(in the infinite top mass limit), including the three-loop matrix elements%
~\cite{Baikov:2009bg,Gehrmann:2010ue,Gehrmann:2010tu},
the one-loop squared single-real-emission contributions%
~\cite{Anastasiou:2013mca,Kilgore:2013gba}
and the convolutions of NNLO cross sections with splitting functions%
~\cite{Hoschele:2012xc,Buehler:2013fha,Hoeschele:2013gga} which require
the knowledge of the NNLO master integrals to higher orders in $\epsilon$%
~\cite{Pak:2011hs,Anastasiou:2012kq}.
Other results are only available as threshold expansions.
They include the partonic cross section of
the purely three-parton real emission~\cite{Anastasiou:2013srw},
the two-loop soft current~\cite{Duhr:2013msa,Li:2013lsa},
the one-loop two emission contribution~\cite{Li:2014bfa},
culminating in the hadronic Higgs production cross section at threshold%
~\cite{Anastasiou:2014vaa}.

In this paper we calculate the dependence of master integrals, appearing
in calculations of the
total inclusive Higgs production cross section via gluon fusion in the infinite top mass limit, on the kinematic
variable $x$, which
is derived by the method of differential equations%
~\cite{Kotikov:1990kg,Kotikov:1991pm,Bern:1993kr,Remiddi:1997ny,Gehrmann:1999as}
(see refs.~\cite{Argeri:2007up,Smirnov:2012gma} for comprehensive reviews).
The differential equations become, however, more and more complicated
with growing loop order.
At \NNNLO{} level
it seems rather difficult to obtain solutions of the differential equations
high enough in the $\epsilon$-expansion
in a naively chosen basis of master integrals.
Recently, a very elegant form of differential equations was introduced
in ref.~\cite{Henn:2013pwa} which is supposed to exist at any loop order.
This conjecture has been strengthened by plenty of examples at two-loop%
~\cite{Henn:2013pwa,Henn:2013woa,Argeri:2014qva,Henn:2014lfa,Caola:2014lpa,Gehrmann:2014bfa} and
three-loop order~\cite{Henn:2013tua,Henn:2013nsa,Caron-Huot:2014lda}
which show the applicability to various kinematic configurations,
even to single-scale integrals~\cite{Henn:2013nsa}.
Although there exist algorithms for constructing an adequate basis in cases of
differential equations depending on $\epsilon$ polynomially%
~\cite{Argeri:2014qva} and
for finite integrals in $D=4$ dimensions~\cite{Caron-Huot:2014lda}
as well as a strategy for the construction from a basis with a triangular
finite part of the homogeneous differential equation matrix%
~\cite{Gehrmann:2014bfa},
a general algorithm to find such a basis is still missing.
However, a lot of methods, tricks and ideas do exist
which are discussed in the references above and used in practice.

The purpose of this paper is twofold.
On the one hand, we review the techniques for finding an adequate basis
using NLO and NNLO master integrals for Higgs production cross section
in sections~\ref{sec:NLO}
and~\ref{sec:NNLO}, respectively, giving the explicit bases as well.
We also present a trick using a characteristic form of higher order
differential equations for the case of coupled master integrals
in section~\ref{subsec:CoupledMasters},
which, to our knowledge, has hitherto not been discussed in the literature.
On the other hand, we show the applicability of the method to
the state of the art problem of finding solutions with full $x$-dependence to
master integrals appearing in \NNNLO{} Higgs production by solving
a non-planar topology in section~\ref{sec:NNNLO}.
In section~\ref{sec:conclusion} we state our conclusions and outlook.

\section{General idea and NLO warm-up}
\label{sec:NLO}

\subsection{Reduction to master integrals}

Suppose that we have families of Feynman integrals, also called topologies, to be evaluated
where the propagator labelled by $i$ is raised to a power $a_i$,
usually called index.
Within dimensional regularization~\cite{'tHooft:1973mm}
integration-by-parts (IBP) identities give linear relations among integrals with
different values of indices $a_i$~\cite{Chetyrkin:1981qh}.
Starting from a large set of values of $a_i$, all integrals can be reduced to
a linearly independent set of master integrals by making use of the IBP identities by means of,
e.g., Laporta algorithm~\cite{Laporta:2001dd}.

We treat phase space integrals contributing to the Higgs production cross
section as cut integrals~\cite{Cutkosky:1960sp}.
In the same way as loop integrals, cut integrals can be reduced to master
integrals via IBP identities by means of the reverse-unitarity method%
~\cite{Anastasiou:2002yz,Anastasiou:2013srw}.
The only difference stems from the fact that integrals containing a cut line with
a non-positive index $a_c \le 0$ vanish.
In order to identify subtopologies, families of Feynman integrals obtained by
setting subsets of indices to be zero,
that have no cuts or are scaleless within dimensional
regularization, we use the private \prog{Mathematica} package \prog{TopoID}.
This code also provides symmetries useful for the reduction
and allows us to identify a minimal set of master integrals.

In this work, we have used an in-house implementation of Laporta algorithm,
as well as the program \prog{FIRE}~\cite{Smirnov:2008iw,Smirnov:2013dia} together with
its unpublished C++ version.
The result is stored in a reduction table for later repeated use.

In the reduction, we use Feynman propagators in Euclidean metric,
which applies also to the master integrals given in this paper.

\subsection{Differential equations for master integrals}

In the case of Higgs production via gluon fusion in the infinite top mass limit,
each topology has only one massive Higgs line and we have forward scattering
kinematics, i.e., the incoming partons' momenta $p_1$ and $p_2$ are equal to
the outgoing partons' momenta $p_3=p_1$ and $p_4=p_2$, respectively.
Therefore, aside from the trivial overall mass scale,
the integrals depend only on one kinematic variable
$x=m_h^2/s$
with $s=(p_1+p_2)^2$ and the space-time dimension $D = 4-2\epsilon$.
Without loss of generality we can set $s=1$.
The derivative of each master integral with respect to $x$ is given, up to
a constant prefactor, by raising the index of the massive line by one
and the resulting integral can be reduced to a linear combination of master integrals.
In this way, we arrive at a set of differential equations for $N$ master
integrals, which can be expressed as the following matrix form:
\begin{align}\label{DEQ}
\partial_x \tilde f(x,\epsilon)= \tilde A(x,\epsilon)\tilde f(x,\epsilon),
\end{align}
where $\tilde f$ is a column vector of master integrals of length $N$
and $\tilde A$ is an $N \times N$ matrix.

\subsection{Change of basis}

The choice of master integrals is not unique and one can always choose
another basis of master integrals.
The basis transformation can be obtained by looking up the entries
in the reduction table for the \emph{new} basis integrals $f$
which are by construction linear combinations of
the \emph{old} basis integrals $\tilde{f}$:
\begin{align}\label{basTransf}
f(x,\epsilon)=B(x,\epsilon)\tilde f(x,\epsilon),
\end{align}
with an $N\times N$ matrix $B$.
Taking the derivative of eq.~\eqref{basTransf} with respect to $x$, one arrives
at the differential equations for the new basis integrals:
\begin{align}\label{NewA}
\partial_x f(x,\epsilon)=A(x,\epsilon) f(x,\epsilon), \qquad
\text{with } A:=\left[(\partial_x B)+B\tilde A\right]B^{-1}.
\end{align}
This means that, providing an alternative basis $f$, we instantly know
the form of its differential equation $A$ by use of the reduction table to
obtain $B$ as well as $\tilde{A}$.

\subsection{Master integral basis in canonical form}

Following Henn's conjecture~\cite{Henn:2013pwa}, a basis $f$ of integrals exists
in which all master integrals become so-called \emph{pure functions}%
\footnote{%
  The number of iterated integrations needed to define a function
  is called weight.
  If a function $f$ consists of terms having a uniform weight
  and if taking a derivative of $f$ also gives a function in which all summands
  have a uniform weight lowered by one,
  then $f$ is called \emph{pure}~\cite{Henn:2013pwa,ArkaniHamed:2010gh}.
  This definition forbids transcendental functions in $f$ from being multiplied
  by algebraic coefficients apart from numbers, thus master integrals given by pure
  functions usually
  have more compact expressions.
}%
and satisfy the differential equations
\begin{align}
 \label{HennDEQ}
  \partial_x f(x,\epsilon)=\epsilon \bar A(x) f(x,\epsilon),
\end{align}
i.e., the dependence of the matrix $A$ on the dimensional parameter $\epsilon$
is factored out as $A=\epsilon \bar A$.
To ensure the basis integrals are pure functions,
$\bar{A}$ should have the form
\begin{equation}
  \label{AbarStruct}
  \bar{A}(x) = \sum_k \frac{\alpha_k}{x-x_k} ,
\end{equation}
where $x_k$ are constants and $\alpha_k$ are constant matrices.
The system of differential equations~\eqref{HennDEQ} can be expanded
in $\epsilon$ as
\begin{equation}
  \partial_x f^{[n]}(x) = \bar{A}(x) f^{[n-1]}(x) , \qquad
  \text{with} \quad f(x,\epsilon) = \sum_{n=-\infty}^{\infty} \epsilon^n f^{[n]}(x) ,
  \label{DEQExp}
\end{equation}
and one can solve it order by order.
The expanded system~\eqref{DEQExp} is triangular in the sense that
only lower order functions $f^{[n-1]}$ appear in the right-hand side
of the set of differential equations for $f^{[n]}$,
hence the solution can be easily obtained in terms of iterated integrals,
provided the boundary condition is fixed at some point $x = x_0$.

The matrix $\bar{A}$ respects singular points of the process, in this
case we have:
for $x=0$ the Higgs line becomes massless and additional infra-red
singularities may be introduced.
In addition, at $x=1$, more precisely for approach from $x<1$, the diagrams develop
a non-zero imaginary part, since the Higgs may be produced indeed.
In our calculation,
we observe another singular point at $x=-1$ for some
NNLO integrals%
\footnote{%
  Our results for the canonical basis integrals contain
  one more singular point for $x \to \infty$ in the unphysical region.
}%
, yielding the canonical form~\eqref{AbarStruct} for the differential
equations:
\begin{align}\label{AForm}
\bar A(x)= \frac{a}{x}+\frac{b}{1-x}+\frac{c}{1+x},
\end{align}
where we find the matrices $a$, $b$ and $c$ to just contain rational numbers.
This assures that at any order the $\epsilon$-expansion of the solution of
the differential equations are iterated integrals expressible
as harmonic polylogarithms (HPLs)~\cite{Remiddi:1999ew},
which can be easily manipulated with \prog{HPL} package implemented in
\prog{Mathematica}%
~\cite{Maitre:2005uu,Maitre:2007kp}.
The first term of the solution in the expansion is a constant,
the next in general contains also HPLs of weight one,
the next in addition HPLs of weight two, etc.
Therefore, the result will be a linear combination of HPLs
with constant prefactors. If the integration constants have suitable weight
the master integrals are pure functions.

\subsection{Parametric representations of integrals}

Although there is no algorithm to obtain an optimal basis from
arbitrary basis integrals in general,
there exist some guiding principles
how to find candidate integrals that may give a canonical form.
For example,
integrals having unit leading singularities%
~\cite{ArkaniHamed:2010gh,Cachazo:2008vp}
are expected to be uniform weight functions.
Another one is investigating parametric representations of integrals,
which is described as follows.

The notion that pure functions are built from iteratively integrated logarithms%
~\cite{Henn:2013pwa}
imposes strong constraints on the candidates.
Sketching the Feynman parameter representation for an integral $I$
(see, e.g., \cite{Smirnov:2012gma})
\begin{align}\label{alphaRep}
I(x,\epsilon)&\sim\int \prod_jd\alpha_j
  \frac{
    \left[U(\{\alpha_i\})\right]^{e_U}
    \prod_i\alpha_i^{a_i-1}
    \delta(\sum_k\alpha_k-1)
  }{
    \left[W(x,\{\alpha_i\})\right]^{e_W}
  },\nonumber\\
e_U&=a-(l+1)D/2,\nonumber\\
e_W&=a-l D/2,\nonumber\\
a&=\sum_i a_i,
\end{align}
where $l$ is the number of loops, $U$ and $W$ are polynomials
in the Feynman parameters $\alpha_i$
and $a_i$ are the corresponding indices.
An integral of form
\begin{align}\label{logForm}
\int \prod_j d\alpha_j \frac{1}{\left[g(\{\alpha_i\},x)\right]^k} ,
\qquad \text{with } k\in \mathbb{N},
\end{align}
where $g$ is an irreducible polynomial, is favored over those of different form
as it yields more likely a pure function in $x$,
see the discussion about d-log forms in ref.~\cite{Henn:2013wfa}.

\begin{figure}[tbp]
  \centering
  \includegraphics[width=.4\textwidth,clip]{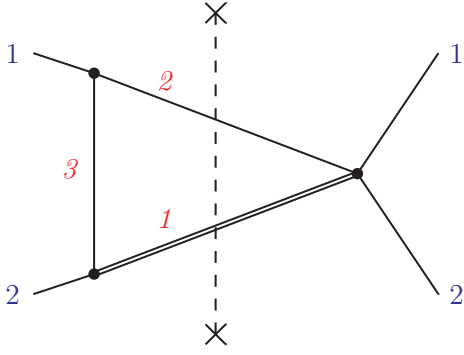}
  \caption{
    \label{fig:NLOtop}
    The NLO topology $\text{TNLO}_2(a_1,a_2,a_3)$.
    The massive Higgs line is depicted by a double line,
    whereas the dashed line denotes the cut.
    Numbers in roman indicate the incoming and outgoing momenta $p_1$ and $p_2$,
    and numbers in italic label the propagators according to the corresponding
    indices.
  }
\end{figure}

Let us illustrate this statement by considering the NLO problem.
After applying symmetries of diagrams
and performing partial fraction decomposition,
one is left with only one topology $\text{TNLO}_2(a_1,a_2,a_3)$ depicted
in fig.~\ref{fig:NLOtop}.
A standard Laporta algorithm finds one master integral,
typically given by TNLO$_2$(1,1,0) obeying the differential equation
\begin{align}
  \partial_x \text{TNLO}_2(1,1,0) = -\frac{1-2\epsilon}{1-x} \text{TNLO}_2(1,1,0).
  \label{TNLO110}
\end{align}
Although the missmatch from eq.~\eqref{HennDEQ} can be cured with a suitable
$x$-normalization (see section~\ref{sec:NNLO}), let us try to understand it
from the parametric representation eq.~\eqref{alphaRep}.
We have $U=1$ from the $\delta$-function for all NLO integrals.
Furthermore, TNLO$_2$(1,1,0) has $a=2$ and therefore $e_W\approx0$,
where we understand the ``$\approx$'' symbol as the $D=4$ approximation%
\footnote{%
  For the purpose of finding candidates in a canonical basis,
  $\epsilon$-dependence of powers can be ignored.
  See also, e.g., ref.~\cite{Henn:2013tua}.
}.
Therefore, we find $k\approx0$ in eq.~\eqref{logForm},
accounting for the non-canonical form of eq.~\eqref{TNLO110}.
The other way around, we need $a=3$ to obtain $k=e_W\approx1$.
This can be achieved, e.g., by raising the index of the massive Higgs line
by one, $\text{TNLO}_2(2,1,0)$, or adding another propagator, $\text{TNLO}_2(1,1,1)$.
In the former case, raising the index of the Higgs line causes an additional $\alpha_1$
in the numerator which cancels against an overall $\alpha_1$ in $W$ of the denominator.
Writing down the differential equations,
we see that the mentioned candidates indeed turn out to form canonical bases:%
\footnote{%
  Although these two integrals obey the same differential equation
  their solutions are different due to different boundary
  conditions.
}%
\begin{align}
\partial_x \text{TNLO}_2(2,1,0) &= \frac{2\epsilon}{1-x} \text{TNLO}_2(2,1,0),\nonumber\\
\partial_x \text{TNLO}_2(1,1,1) &= \frac{2\epsilon}{1-x} \text{TNLO}_2(1,1,1) .
\end{align}

It is important to remember this fact in the following
since these diagrams will appear as subgraphs at higher loop order.
Performing the same manipulations, i.e. raising one index of a bubble or stretching it into
a triangle, for the subgraphs will lead to promising candidates
(see ref.~\cite{Henn:2013tua} as well).
In general, we observe that there are cases where adding additional lines
or raising indices helps.
Finally, it is worth mentioning that the arguments given here
have made use of the diagrammatic structure of the integrals,
but (apart from the $\alpha_1$ cancellation mentioned above) not of the explicit
structure of the $W$ polynomial in eq.~\eqref{alphaRep} and
therefore were (almost) independent of the kinematics.
Hence, it is not surprising that some of the integrals in canonical bases given
in section~\ref{sec:NNLO} and section~\ref{sec:NNNLO} resemble results
for similar topologies
with different kinematics found in the literature.

\section{NNLO: examples and solutions}
\label{sec:NNLO}

\subsection{Known techniques}
\label{subsec:KnownTechniques}

\begin{figure}[tbp]
  \centering
  \setlength{\tabcolsep}{6pt}
  \renewcommand{\arraystretch}{6}
  \renewcommand{\GraphSize}{.3}
  \begin{tabular}{ccc}
    \IncGraph{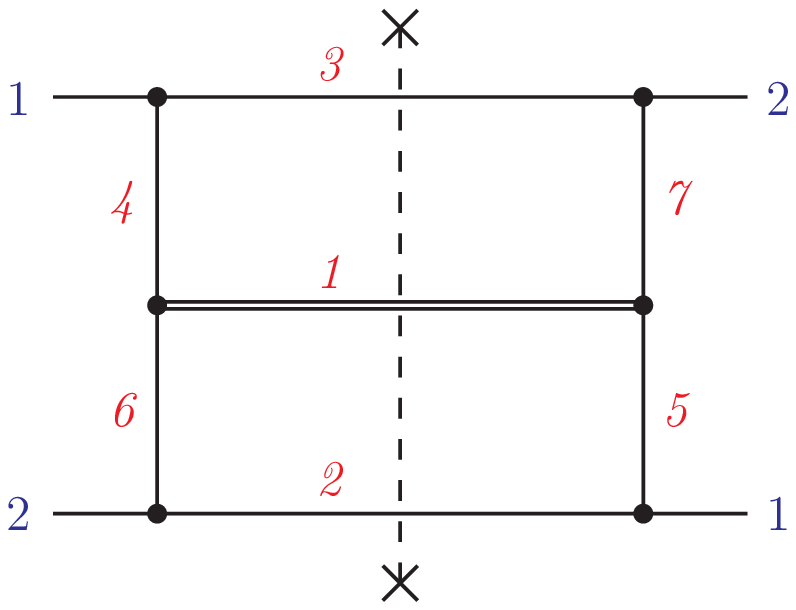} &
    \IncGraph{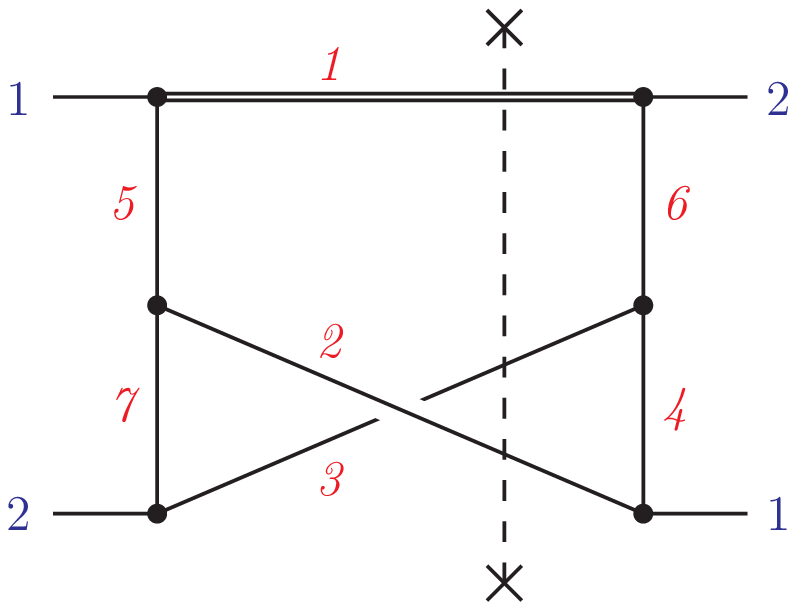} &
    \IncGraph{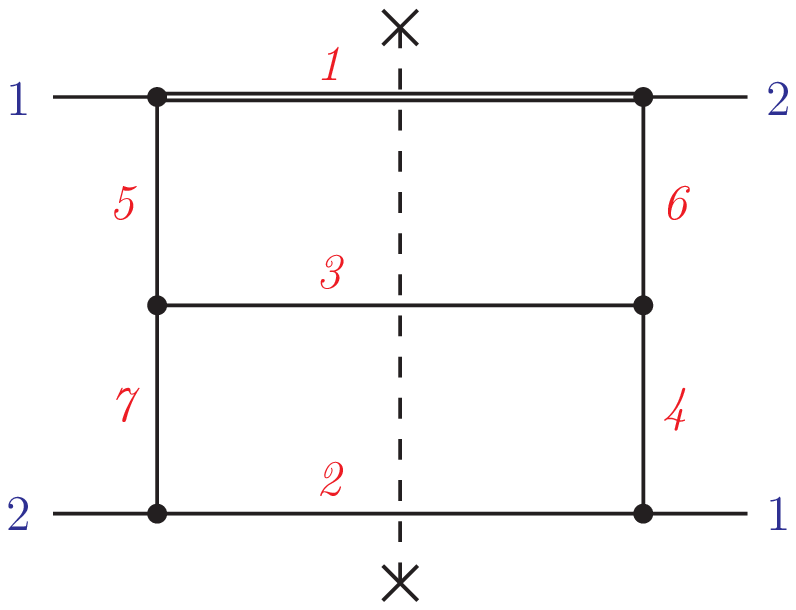} \\
    \IncGraph{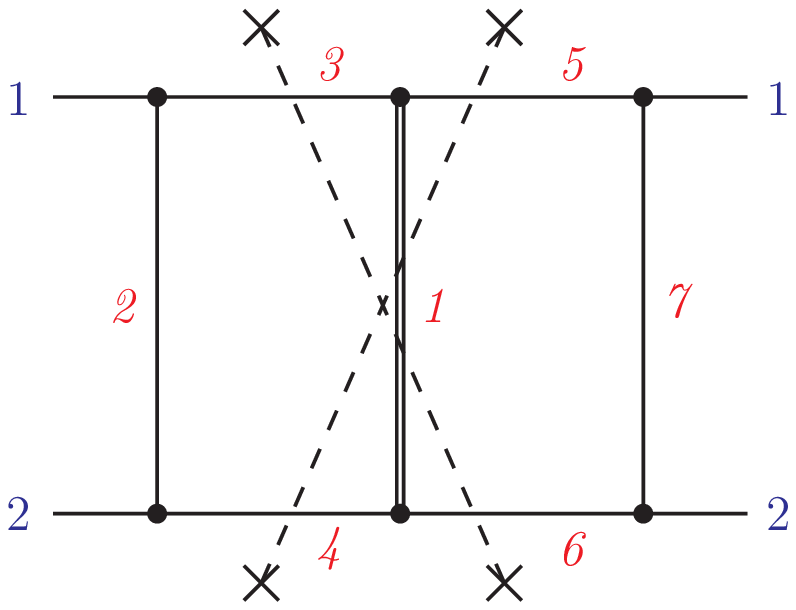} &
    \IncGraph{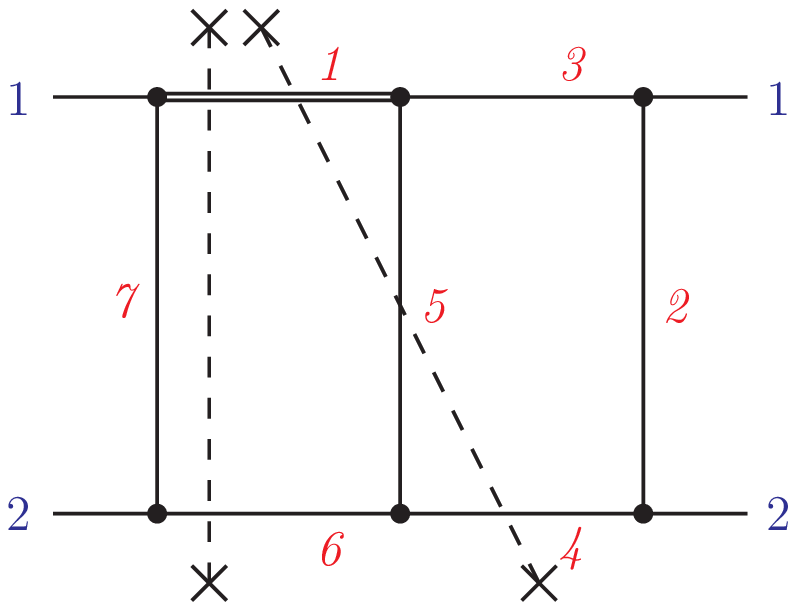} &
    \IncGraph{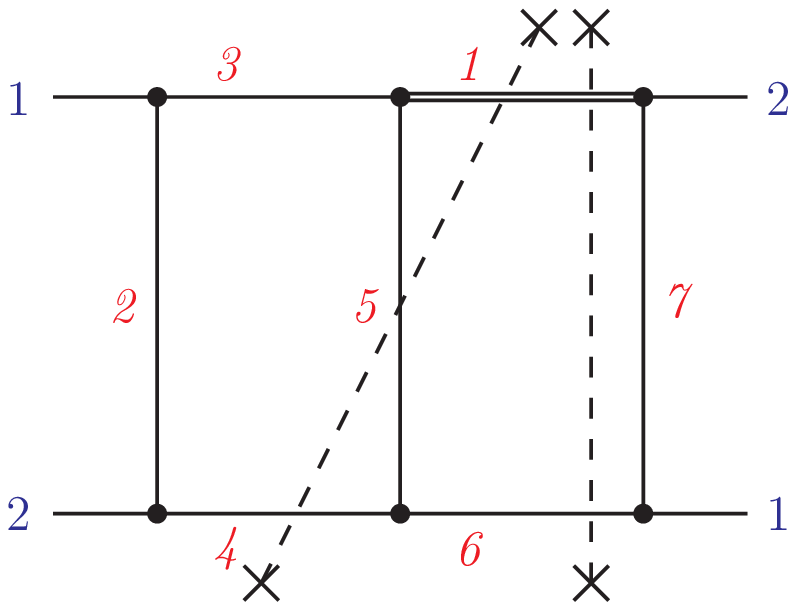} \\
    \IncGraph{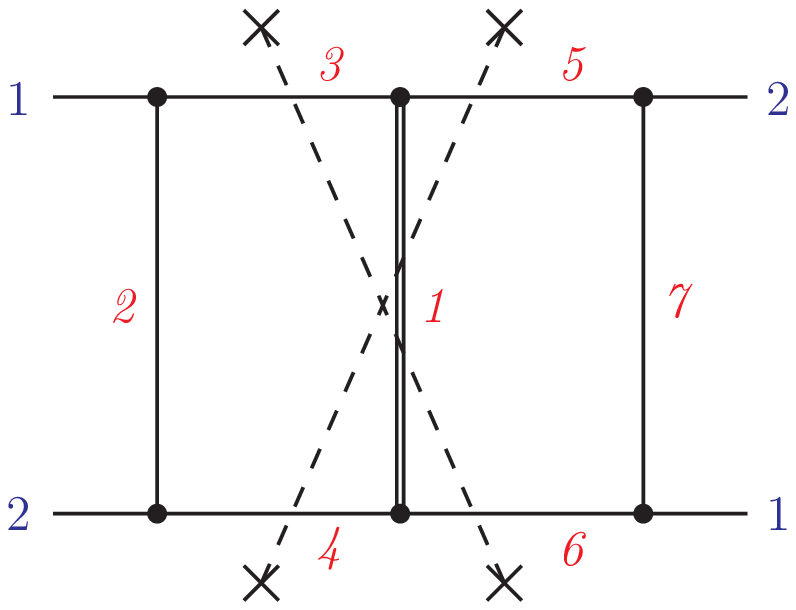} &
    \IncGraph{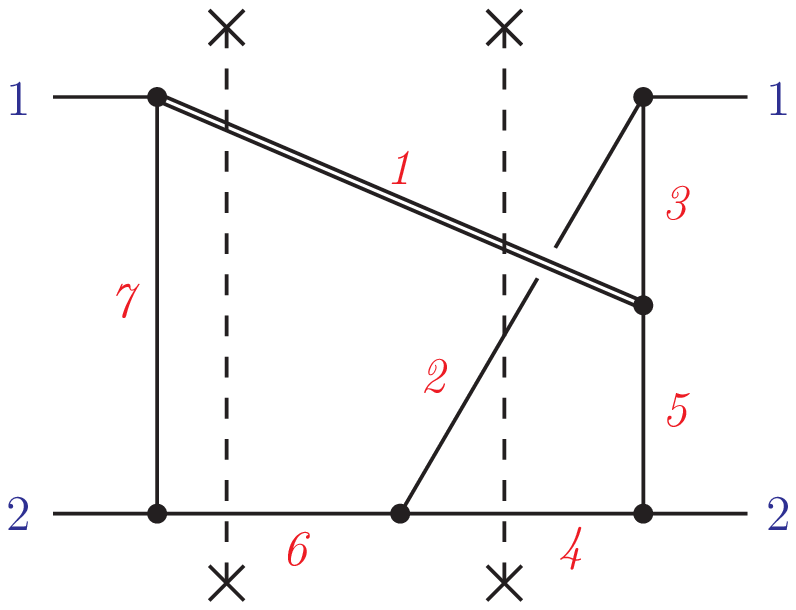} &
    \IncGraph{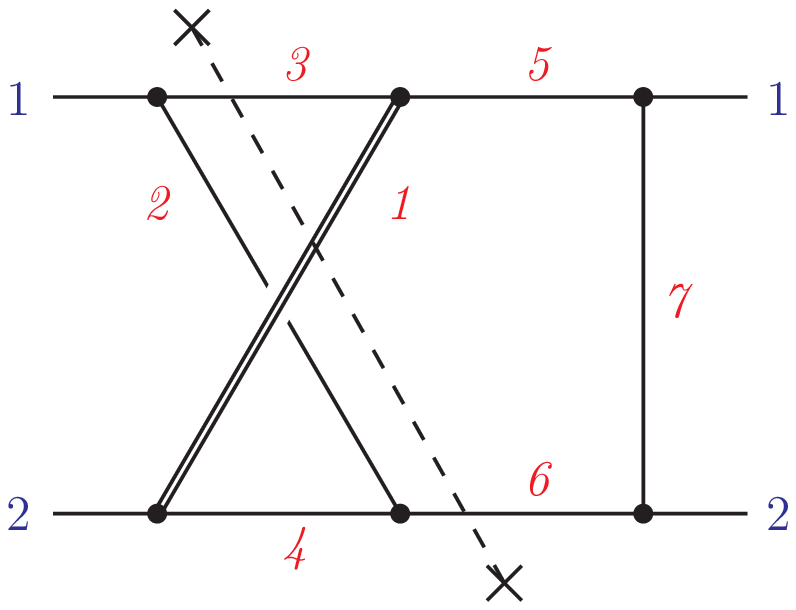}
  \end{tabular}
  \caption{
    \label{fig:NNLOtops}
    The NNLO topologies $\text{TT}X_c(a_1,a_2,a_3,a_4,a_5,a_6,a_7)$
    involving our choice for a canonical basis.
    The subscript $c = 2, 3$ of topologies distinguishes two-particle cuts and
    three-particle cuts.
    The massive Higgs line is depicted by a double line,
    whereas dashed lines denote possible cuts.
    Numbers in roman indicate the incoming and outgoing momenta $p_1$ and $p_2$
    and numbers in italic label the propagators according to the corresponding
    indices.
    In the text we define all integrals as single-cut integrals.
    For TTE and TTH, two cuts give the same contribution but only one of them
    is taken into account in the defnition of the corresponding master integrals.
  }
\end{figure}

Let us discuss further known tricks for finding an adequate basis
by looking at the example of the three-particle phase space
diagram defined in terms of the topology $\text{TTA}_3$
(see fig.~\ref{fig:NNLOtops}) occurring at NNLO via
\begin{align}
\text{TTA}_3(1,1,1,0,0,0,0)=:\text{TTA}_3(1,1,1),
\end{align}
where we omit the trailing zeros in the indices for simplicity.
In the reduction basis obtained from our reduction table,
its differential equation is coupled to another integral
$\text{TTA}_3(1,1,1,-1,0,0,0)$ having an additional scalar product
in the numerator.
For the purpose of finding good candidates we raise one index of the massless bubble%
~\cite{Henn:2013tua}.
It is well known that integrating a massless bubble with the indices
$b_1$ and $b_2$ gives, up to a prefactor,
a propagator with the index $b_1+b_2-2+\epsilon$.
Therefore, we have
\begin{align}
V_1^*=\text{TTA}_3(2,2,1)\sim\text{TNLO}_2(2,1+\epsilon,0)
\end{align}
and we expect this to be a good candidate from the discussion
for the NLO case in section~\ref{sec:NLO}.
The second candidate which couples to this one can be found by constructing
a subtle \emph{linear combination}.
For that purpose,
let us compare eq.~\eqref{alphaRep} to eq.~\eqref{logForm} for $k=1$,
i.e. in a first step we set $e_W\approx k=1$.
This means that we have fixed $a=5$ and $e_U\approx-1$.
For example the parametric representation of the above candidate $V_1^*$
is of the type
\begin{align}\label{alphaRepPS3}
\text{TTA}_3(2,2,1)&\sim
  \int \prod_jd\alpha_j\frac{\alpha_1\alpha_2\delta(\sum_k\alpha_k-1)}{
    \left(\alpha_1\alpha_2+\alpha_1\alpha_3+\alpha_2\alpha_3\right)W(x,\{\alpha_i\})}.
\end{align}
One could try $\text{TTA}_3(1,2,2)$ which is of the same form,
but with $\alpha_1\alpha_2$ in the numerator replaced by $\alpha_2\alpha_3$.
However, the candidate does not lead to a differential equation of
the desired form.
Therefore, one can try to identify $W$ in eq.~\eqref{alphaRepPS3}
with $g$ in eq.~\eqref{logForm}, which means we have to cancel
the $U$ polynomial $(\alpha_1 \alpha_2 + \alpha_1\alpha_3 + \alpha_2\alpha_3)$.
This is achieved by adding three integrals to
\begin{align}
V_2^*
  =\text{TTA}_3(2,2,1)+\text{TTA}_3(2,1,2)+\text{TTA}_3(1,2,2)
  =2\text{TTA}_3(2,2,1)+\text{TTA}_3(1,2,2).
\end{align}
Note that the construction given above can be generalized to any loop order
for this type of sunrise diagrams
(see the results of section~\ref{sec:NNNLO} for the three-loop case).

Constructing the differential equations for the candidates
$V_1^*$ and $V_2^*$ using eq.~\eqref{NewA}
we find only $A_{22}$ to be of inappropriate form
\begin{align}
\lim_{\epsilon\to0} A =
\begin{pmatrix}
 0 & 0\\
 0 & \frac{1}{1-x}
\end{pmatrix}.
\end{align}
The non-vanishing \emph{diagonal}
element corresponds to the homogeneous differential
equation of $V_2^*$ in lowest order in $\epsilon$
and therefore it appears in all orders~\cite{Argeri:2007up}.
The problem is resolved by allowing for an $\epsilon$ independent
\emph{$n(x)$-normalization}, i.e.
a shift $V_2^*\to n(x)V_2^*$, yielding
\begin{align}
\lim_{\epsilon\to0}A_{22}=\frac{1}{1-x}+\frac{\partial_x n(x)}{n(x)} \overset{!}{=}0.
\end{align}
This determines $n(x)=1-x$ and we have found two elements of the canonical basis
\begin{align}\label{V1V2}
V_1&=\epsilon\text{TTA}_3(2,2,1),\nonumber\\
V_2&=\epsilon(1-x)\bigl[2\text{TTA}_3(2,2,1)+\text{TTA}_3(1,2,2)\bigr],
\end{align}
obeying the differential equation with
\begin{equation}
  \setlength{\arraycolsep}{6pt}
  \label{A2Ints}
  A = \epsilon
  \begin{pmatrix}
    -\frac{3}{x} & \frac{1}{x} + \frac{1}{1-x} \\
    -\frac{6}{x} & \frac{2}{x} + \frac{4}{1-x}
  \end{pmatrix} .
\end{equation}
The prefactors of $\epsilon$ have been introduced to make the integrals start
at finite order.

Note that the result, we have found here, will not be changed
by adding more master integrals to the considerations.
In this sense, finding a canonical basis can be approached \emph{step by step},
starting with the integrals with lowest number of lines and continuously
increasing that number.
Master integrals with coupled differential equations have to be added in one step to
the problem, but here other strategies apply, as we show
in section.~\ref{subsec:CoupledMasters}.

Now we add the next master integral of the topology $\text{TTA}_3$
to the problem, given by
$V_3^*=\text{TTA}_3(1, 1, 1, 1, 0, 1, 0)$,
where we have just taken the integral from the reduction basis as a candidate.
Since it has $a=5$ in its parametric representation~\eqref{alphaRep},
this is a good choice as motivated above.
The matrix defining the differential equations becomes
\begin{align}\label{A3Ints}
\setlength{\arraycolsep}{6pt}
A =
\begin{pmatrix}
 -\frac{3\epsilon}{x} & \frac{\epsilon}{x} + \frac{\epsilon}{1-x} & 0\\
 -\frac{6 \epsilon}{x} & \frac{2 \epsilon}{x} + \frac{4 \epsilon}{1-x} & 0\\
-\frac{1}{\epsilon^2 x} & \frac{1}{2\epsilon^2 x} & -\frac{2\epsilon}{x}
\end{pmatrix},
\end{align}
where the upper left $2\times 2$ block corresponds to $(V_1, V_2)$ and
forms a canonical basis already as eq.~\eqref{A2Ints}.
The \emph{off-diagonal} elements in the last row depend on $\epsilon$
differently from the desired form.
These elements correspond to the inhomogeneous terms in the differential equation
for $V_3^*$.
The problem is cured by \emph{$n(\epsilon)$-normalization}, i.e.
by changing $V_3^*\to n(\epsilon)V_3^*$.
In this case we find $n(\epsilon)=\epsilon^3$, such that
\begin{align}
V_3=\epsilon^3\text{TTA}_3(1, 1, 1, 1, 0, 1, 0)
\end{align}
completes a canonical basis of the $3\times 3$ subproblem.

In summary, a diagonal scaling matrix
\begin{align}
  \label{scaleB}
  B_{ij}^\text{s} = \begin{cases} s_i(x,\epsilon), & i = j, \\ 0, & i \ne j , \end{cases}
\end{align}
changes the coefficient matrix as
\begin{align}
  \label{scaleA}
  A_{ij}^\text{s} &= \frac{\partial_x s_i}{s_i} \delta_{ij}
    + \frac{s_i}{s_j} A_{ij}.
\end{align}

Lastly, let us remark that bases exist where $\epsilon$ factorizes from their
$A$ matrix as in eq.~\eqref{HennDEQ} but the matrix is not of the
desired form given in eq.~\eqref{AForm}.
For example, choosing
\begin{align}
V_3^*= (1 + \epsilon)\epsilon^2 x \text{TTA}_3(1, 1, 1, 2, 0, 1, 0)
\end{align}
instead of $V_3$ the last row in eq.~\eqref{A3Ints} changes to
\begin{align}
\setlength{\arraycolsep}{6pt}
A =
\begin{pmatrix}
 -\frac{3\epsilon}{x} & \frac{\epsilon}{x} + \frac{\epsilon}{1-x} & 0\\
 -\frac{6 \epsilon}{x} & \frac{2 \epsilon}{x} + \frac{4 \epsilon}{1-x} & 0\\
 -3 \epsilon & \frac{\epsilon}{2}+\frac{\epsilon}{1-x} & -\frac{2 \epsilon}{x} \\
\end{pmatrix},
\end{align}
where the off-diagonal elements in the last row induce non-logarithmic functions
in the solution of $V_3^*$ and therefore the result cannot be
a pure function.

\subsection{Techniques for coupled master integrals}
\label{subsec:CoupledMasters}

The techniques we want to discuss next touches on the issue of master integrals
coupled by their differential equations.
In particular, a problem that occurs frequently is the following:
when there is a system of $n$ coupled master integrals in a reduction basis,
one tries a set of $n$ candidates for a canonical basis and sees
if the resulting $n$ differential equations are of the canonical form or not.
Even if one of the $n$ candidates is a \emph{good} integral that would form
a canonical basis with appropriately chosen other $(n-1)$ good integrals,
the corresponding differential equation may not be of the canonical form
due to a choice of the other $(n-1)$ \emph{bad} integrals, which makes it
difficult to identify good integrals.
Nonetheless, it would be worthwhile knowing
if one of the $n$ candidates is a canonical basis integral
so one could keep good integrals and dismiss bad integrals.

In the following we want to show how to distinguish suitable candidates for
canonical master integrals from unsuitable choices using the fact that
the system of $n$ coupled first order
differential equations is equivalent to one $n$th order differential equation
for one of the master integrals.
The idea is that the resulting higher order differential equation is unique
for each master integral in the sense that it is independent of eliminated integrals.
It defines the master integral itself and
furthermore takes a specific form for canonical master integrals, founding on
the assumption that a set of canonical master integrals exists.
Moreover, we will show that once one of canonical master integrals in a
coupled system is found the assumption of the existence of a canonical basis
allows us to construct a set of the other canonical master integrals coupled
to it.

\subsubsection{Characteristic form of higher order differential equations}

Let us discuss the situation of two coupled master integrals $f_1$ and $f_2$
to explain the method in detail:
\begin{align}
  \label{CoupSystOf2}
  f_1'&=a_{11}f_1+a_{12}f_2+\sum_i r_{1i} g_i,\nonumber\\
  f_2'&=a_{21}f_1+a_{22}f_2+\sum_i r_{2i} g_i,
\end{align}
where primes denote derivatives with respect to $x$ and
$g_i$ in the right-hand sides are master integrals assumed to be fixed already
and to form a canonical basis, obeying
\begin{align}
  \label{DEQrhs}
  g_i' = \sum_j \alpha_{ij} g_j .
\end{align}
All the quantities given here depend on $x$ and $\epsilon$.
Taking one more derivative with respect to $x$ of the first line of
eq.~\eqref{CoupSystOf2} we find
\begin{align}
  f_1'' =
    a_{11}' f_1 + a_{11} f_1' + a_{12}' f_2 + a_{12} f_2'
    + \sum_i \left( r_{1i}' g_i + r_{1i} g_i' \right) .
\end{align}
Eliminating $f_2$ and $f_2'$ by eq.~\eqref{CoupSystOf2} and $g_i'$
by eq.~\eqref{DEQrhs} we obtain a second order differential equation for $f_1$:
\begin{align}
  f_1'' &=
    - \left( - a_{11} - \frac{a_{12}'}{a_{12}} - a_{22} \right) f_1'
    + \left( a_{11}' - \frac{a_{11} a_{12}'}{a_{12}} + a_{12} a_{21}
             - a_{11} a_{22} \right) f_1
  \nonumber \\ & \quad
    + \sum_i \biggl( - \frac{a_{12}' r_{1i}}{a_{12}} - a_{22} r_{1i}
                     + a_{12} r_{2i} + r_{1i}' + \sum_j r_{1j} \alpha_{ji}
             \biggr) g_i
  \nonumber \\ &
    =: - C_1 f_1' + C_0 f_1 + \sum_i C_{0i} g_i .
  \label{SecondOrderDiffGen}
\end{align}
It is important to emphasize that this differential equation is independent of
$f_2$ and uniquely defines the behaviour of $f_1$.
The coefficients $C_1$, $C_0$ and $C_{0i}$ are invariant under any basis
transformations that change only $f_2$ as
$f_2 \to b_{21} f_1 + b_{22} f_2 + \sum_i \beta_{2i} g_i$.

\paragraph{Case 1: $f_1$ and $f_2$ are canonical master integrals.}
Let us now reveal what characteristic form for the higher order differential
equation of $f_1$ should appear when $f_1$ and $f_2$ are canonical master
integrals as $g_i$ are.
In such a basis, $a_{ij}$ and $r_{ij}$ as well as $\alpha_{ij}$ are proportional
to $\epsilon$.
Therefore, the coefficients $C_1$, $C_0$ and $C_{0i}$ defined by
eq.~\eqref{SecondOrderDiffGen} can be decomposed into $\epsilon$ independent
coefficients as
\begin{align}
  \label{Cs}
  C_1 &= C_1^{(0)} + \epsilon C_1^{(1)}, \nonumber \\
  C_0 &= \epsilon C_0^{(1)} + \epsilon^2 C_0^{(2)}, \nonumber \\
  C_{0i} &= \epsilon C_{0i}^{(1)} + \epsilon^2 C_{0i}^{(2)},
\end{align}
given by
\begin{align}
  \label{coefficientsCs}
  C_1^{(0)} &= - \frac{a_{12}'}{a_{12}}, \nonumber \\
  C_1^{(1)} &= \frac{1}{\epsilon}\left(-a_{11}-a_{22}\right), \nonumber \\
  C_0^{(1)} &= \frac{1}{\epsilon}\left(a_{11}'-\frac{a_{11}a_{12}'}{a_{12}}
               \right), \nonumber \\
  C_0^{(2)} &= \frac{1}{\epsilon^2}\left(a_{12}a_{21}-a_{11}a_{22}
               \right), \nonumber \\
  C_{0i}^{(1)} &= \frac{1}{\epsilon}\left(-\frac{a_{12}' r_{1i}}{a_{12}}
                  +r_{1i}'\right), \nonumber \\
  C_{0i}^{(2)} &= \frac{1}{\epsilon^2}\biggl(-a_{22}r_{1i}+a_{12}r_{2i}
                  +\sum_j r_{1j}\alpha_{ji}\biggr).
\end{align}

\paragraph{Case 2: $f_1$ is a canonical master integral but $f_2$ is not.}
Even if $f_2$ is not a canonical master integral and it makes
$a_{ij}$ and $r_{ij}$ not be of canonical form,
we can utilize the uniqueness of the coefficients $C_1$, $C_0$ and $C_{0i}$
in the higher order differential equation for $f_1$.
They must still have decompositions in $\epsilon$ like eq.~\eqref{Cs},
although the identities for $C_1^{(m)}$, $C_0^{(m)}$, $C_{0i}^{(m)}$
in terms of $a_{ij}$ and $r_{ij}$ eq.~\eqref{coefficientsCs} do not hold
any longer.
Furthermore, it allows us to reconstruct what coefficients $a_{ij}^\text{p}$
and $r_{ij}^\text{p}$ in a system of differential equations would be within
a basis in which $f_2$ is properly chosen to be a canonical master integral
$f_2^\text{p}$, under the assumption that such an $f_2^\text{p}$ does exist.
Since in such a proper basis $C_1$, $C_0$ and $C_{0i}$ take the same form as
\textbf{Case 1} in terms of $a_{ij}^\text{p}$ and $r_{ij}^\text{p}$,
we can invert eq.~\eqref{coefficientsCs} to obtain
$a_{ij}^\text{p}$ and $r_{ij}^\text{p}$:
\begin{align}
  \label{AsFromCs}
  {a_{12}^\text{p}}'+C_1^{(0)}a_{12}^\text{p} &= 0, \nonumber\\
  {a_{11}^\text{p}}'-\frac{{a_{12}^\text{p}}'}{a_{12}^\text{p}}a_{11}^\text{p}
    &= \epsilon C_0^{(1)}, \nonumber\\
  a_{22}^\text{p} &= -a_{11}^\text{p}-\epsilon C_1^{(1)}, \nonumber\\
  a_{21}^\text{p} &= \frac{a_{11}^\text{p}a_{22}^\text{p}}{a_{12}^\text{p}}
    +\epsilon^2 \frac{C_0^{(2)}}{a_{12}^\text{p}}, \nonumber\\
  {r_{1i}^\text{p}}'-\frac{{a_{12}^\text{p}}'}{a_{12}^\text{p}}r_{1i}^\text{p}
    &= \epsilon C_{0i}^{(1)}, \nonumber\\
  r_{2i}^\text{p} &= \frac{a_{22}^\text{p}r_{1i}^\text{p}}{a_{12}^\text{p}}
    -\sum_j\frac{r_{1j}^\text{p}\alpha_{ji}}{a_{12}^\text{p}}
    +\epsilon^2 \frac{C_{0i}^{(2)}}{a_{12}^\text{p}}.
\end{align}
We can solve this system of differential equations, line by line for $a_{ij}$
and $r_{ij}$, setting the integration constants of $a_{12}$, $a_{11}$ and
$r_{1i}$ to constants proportional to $\epsilon$, namely $\epsilon c_{12}$,
$\epsilon c_{11}$ and $\epsilon k_{1i}$, respectively.

\subsubsection{Construction of canonical basis}
\label{subsubsec:B-construction}

Having all entries of $a_{ij}^\text{p}$ and $r_{ij}^\text{p}$
as in \textbf{Case 2} of the previous section,
one can construct a canonical master integral $f_2^\text{p}$ that satisfies
the differential equations implied by % eq.~\eqref{AsFromCs}
$a_{ij}^\text{p}$ and $r_{ij}^\text{p}$.
The linear basis transformation
\begin{align}\label{Bc2Constr}
B =
\begin{pmatrix}
 \mathbbm{1} & 0 & 0\\
 0 & 1 & 0\\
 \beta_{2i} & b_{21} & b_{22}
\end{pmatrix}
\end{align}
from the basis $(g_i, f_1, f_2)$ obeying the differential equation with
the matrix $A$
to the canonical basis $(g_i, f_1, f_2^\text{p})$ with $A^\text{p}$
satisfies eq.~\eqref{NewA}, or
\begin{align}
  \label{deqForBS}
  B' = A^\text{p} B - B A,
\end{align}
where the matrix $A$ and $A^\text{p}$ are given by
\begin{equation}
  A = \begin{pmatrix}
        \alpha_{ij} & 0 & 0 \\
        r_{1i} & a_{11} & a_{12} \\
        r_{2i} & a_{21} & a_{22}
      \end{pmatrix} , \qquad
  A^\text{p} = \begin{pmatrix}
        \alpha_{ij} & 0 & 0 \\
        r_{1i}^\text{p} & a_{11}^\text{p} & a_{12}^\text{p} \\
        r_{2i}^\text{p} & a_{21}^\text{p} & a_{22}^\text{p}
      \end{pmatrix} .
\end{equation}
Each component in the row corresponding to $f_1$ (the next to the last line)
of eq.~\eqref{deqForBS}
gives a linear equation for $\beta_{2i}$, $b_{21}$ and $b_{22}$, respectively.
The rows above the mentioned one give trivial equations,
whereas the row below gives differential equations that
serve as consistency checks.
Once the basis transformation $B$ is determined, one can obtain an explicit
expression of $f_2^\text{p}$ as a linear combination of $g_i$, $f_1$ and $f_2$.

Note that until the end we do not need to fix the integration constants
$c_{12}$, $c_{11}$ and $k_{1i}$ introduced in \textbf{Case 2}.
Since from eq.~\eqref{AsFromCs} $a_{12}^\text{p}$ is proportional to $c_{12}$
and $a_{21}^\text{p}$ and $r_{2i}^\text{p}$ are proportional to $c_{12}^{-1}$
whereas $a_{11}^\text{p}$, $a_{22}^\text{p}$ and $r_{1i}^\text{p}$ are
independent of it,
$c_{12}^{-1}$ can be interpreted as a numerical normalization factor of
$f_2^\text{p}$, see eqs.~\eqref{scaleB} and~\eqref{scaleA}.
In addition to the normalization factor $c_{12}^{-1}$,
$c_{11}$ and $k_{1i}$ span a multi-dimensional space of solutions
for $f_2^\text{p}$ and cover
the full class of canonical master integrals that are coupled partners to $f_1$.

So far, we have seen that if $f_1$ is a canonical master integral
one can construct another canonical master integral $f_2^\text{p}$
coupled to $f_1$.
This leads to an algorithm to see whether $f_1$ can be a canonical master integral:
assuming $f_1$ is a canonical master integral, one tries to construct
$f_2^\text{p}$ on the basis of the above considerations.
If it fails at any step, one concludes that $f_1$ cannot be a canonical master
integral. Once $f_2^\text{p}$ is explicitly constructed, one can see whether
$f_1$ and $f_2^\text{p}$ form a canonical basis as they should, which is
equivalent to find a consistent solution of $B$ in eq.~\eqref{deqForBS}.
The details are as follows:
\begin{enumerate}
  \item Calculate the coefficients $C_1$, $C_0$ and $C_{0i}$ in the higher order
        differential equation for $f_1$ from $a_{ij}$ and $r_{ij}$ by
        eq.~\eqref{SecondOrderDiffGen}.
  \item Assuming $f_1$ is a canonical master integral,
        one should find that $C_1$, $C_0$ and $C_{0i}$
        have decompositions in $\epsilon$ as eq.~\eqref{Cs},
        otherwise $f_1$ cannot be a canonical master integral
        and should be dismissed.
  \item Reconstruct
        $a_{ij}^\text{p}$ and $r_{ij}^\text{p}$ from $C_1^{(m)}$, $C_0^{(m)}$
        and $C_{0i}^{(m)}$
        by eq.~\eqref{AsFromCs}.
        If one requires them to be of the desired form
        \begin{equation}
          \label{aijHennmasters}
          a_{ij}^\text{p}, r_{ij}^\text{p} \sim
          \epsilon \left( \frac{n_0}{x} + \frac{n_1}{1-x} + \frac{n_{-1}}{1+x}
          \right) ,
        \end{equation}
        where $n_0$, $n_1$ and $n_{-1}$ are numbers,
        the differential equations for $a_{12}^\text{p}$, $a_{11}^\text{p}$
        and $r_{1i}^\text{p}$ must be easily solved, and if they are difficult
        to solve most likely they do not have the above form%
        \footnote{%
          Actually, eq.~\eqref{aijHennmasters} can be taken as an ansatz for
          the differential equations such that all differential equations
          we need to solve are reduced to algebraic equations.
        }%
        .
        If one of the reconstructed entries $a_{ij}^\text{p}$ and $r_{ij}^\text{p}$ are not
        of the form eq.~\eqref{aijHennmasters}, $f_1$ cannot be a canonical master integral.
  \item Find the basis transformation~\eqref{Bc2Constr}
        to the basis that satisfies the differential equations given by
        $a_{ij}^\text{p}$ and $r_{ij}^\text{p}$ from eq.~\eqref{deqForBS}.
        From this transformation one obtains $f_2^\text{p}$.
        If there is no solution of eq.~\eqref{deqForBS} with eq.~\eqref{Bc2Constr},
        $f_1$ cannot be a canonical master integral.
\end{enumerate}

Note that in our case eq.~\eqref{aijHennmasters} is imposed
on the canonical form,
but for the derivation of $f_2^\text{p}$ we only needed the assumption that
$a_{ij}^\text{p}$ and $r_{ij}^\text{p}$ are proportional to $\epsilon$.
Therefore this algorithm should also work for other calculations within the framework
of canonical differential equations having different forms~%
\eqref{AbarStruct} in $x$.

\subsubsection{Example in the three-particle phase space at NNLO}

Let us apply the above algorithm to the example of the two coupled
phase space integrals of
$\text{TTA}_3(a_1, a_2, a_3)$ we have already encountered
in section~\ref{subsec:KnownTechniques}.
More specifically, we put
\begin{align}\label{coupledExampleNNLO}
f_1&=\epsilon \text{TTA}_3(2,2,1),\nonumber\\
f_2&=\text{TTA}_3(2,1,1).
\end{align}
We have previously seen that $f_1 = V_1$ is a canonical master integral,
nevertheless we will apply the algorithm to this pair of integrals and
see what happens.
The differential equations have no inhomogeneous terms and therefore
\begin{align}
r_{1i}&=0,\nonumber\\
r_{2i}&=0.
\end{align}
Writing down the second order differential equation for $f_1$,
we can reconstruct $A^\text{p}$ from its coefficients:
\begin{align}
  \setlength{\arraycolsep}{6pt}
  A^\text{p}=\epsilon
  \begin{pmatrix}
    \frac{c_{11}}{x} + \frac{c_{11}+3}{1-x} &
    \frac{c_{12}}{x} + \frac{c_{12}}{1-x} \\
    - \frac{c_{11} (c_{11}+1)}{c_{12} x}
      - \frac{(c_{11}+3)(c_{11}-1)}{c_{12} (1-x)} &
    - \frac{c_{11}+1}{x} - \frac{c_{11}-1}{1-x}
  \end{pmatrix} .
\end{align}
Correspondingly, we can find $f_2^\text{p}$ as
\begin{equation}
  f_2^\text{p}
  =
  \frac{1-c_{11}(1-4\epsilon)-2\epsilon(1+x)}{c_{12}(1-4\epsilon)} f_1
  - \frac{2\epsilon(1-2\epsilon)(1-3\epsilon)}{c_{12}(1-4\epsilon)} f_2 .
\end{equation}
Once $f_2^\text{p}$ is obtained, one can easily check that $f_1$ and
$f_2^\text{p}$ form a canonical basis with $A^\text{p}$.

Note that we have not fixed the integration constants $c_{11}$ and $c_{12}$.
As discussed before, $c_{12}$ determines the normalization of $f_2^\text{p}$.
If we choose $c_{11}=-3$ and $c_{12}=1$, $A^\text{p}$ turns into
eq.~\eqref{A2Ints} and $f_2^\text{p}$ becomes $V_2$, which can be verified
by the reduction.
Other interesting solutions are given by $c_{11}=1$, $c_{12}=1$ or
$c_{11}=-1$, $c_{12}=2$, for which $a_{21}^\text{p}=a_{22}^\text{p}$
and the second row of $A^\text{p}$ becomes independent of $1/(1-x)$
or $1/x$, respectively.

By contrast, if we interchange $f_1$ and $f_2$ in the above example%
~\eqref{coupledExampleNNLO} and put
\begin{align}
f_1&=\text{TTA}_3(2,1,1),\nonumber\\
f_2&=\epsilon\text{TTA}_3(2,2,1) ,
\end{align}
the first coefficient stemming from $C_1^{(0)}$ already yields
\begin{align}
a_{12}^\text{p}&=-\epsilon c_{12} + \epsilon\frac{c_{12}}{x} ,
\end{align}
in disagreement with the desired form~\eqref{aijHennmasters}.
Therefore we can conclude $f_1=\text{TTA}_3(2,1,1)$ cannot be
a canonical master integral.

\subsubsection{Example with a non-zero inhomogeneity}

As we will see in section~\ref{subsec:NNLOsol},
there is another pair of coupled integrals $\text{V}_{11}$ and
$\text{V}_{12}$ among the NNLO canonical master integrals.
Their differential equations contain $V_1$ and $V_2$ as
inhomogeneous terms.
Let us apply the algorithm to a basis in which $V_{11}$ is correctly chosen
as well as $V_1$ and $V_2$ but the coupled integral is chosen differently from
$V_{12}$.
By putting
\begin{align}
 g_1 &= V_{1\phantom{1}} = \epsilon \text{TTA}_3(2,2,1), \nonumber\\
 g_2 &= V_{2\phantom{1}} = \epsilon (1-x) \left[\text{TTA}_3(1,2,2)+2 \text{TTA}_3(2,2,1)\right], \nonumber\\
 f_1 &= V_{11} = \epsilon^3 \text{TTF}_3(1,0,1,1,1,0,1), \nonumber\\
 f_2 &= \epsilon^2 \text{TTF}_3(1, 0, 2, 1, 1, 0, 1) ,
\end{align}
the algorithm gives $A^\text{p}$ in the desired form:
\begin{equation}
  A^\text{p} =
  \epsilon \left( \frac{a^\text{p}}{x} + \frac{b^\text{p}}{1-x} \right) ,
\end{equation}
with
\begin{equation}
  \setlength{\arraycolsep}{6pt}
  a^\text{p} =
    \begin{pmatrix}
      -3 & 1 & 0 & 0 \\
      -6 & 2 & 0 & 0 \\
      k_{11} & k_{12} & c_{11} & c_{12} \\
      - \frac{c_{11} k_{11} - 6 k_{12} + 2}{c_{12}}
      & - \frac{2 c_{11} k_{12} + 2 k_{11} + 10 k_{12} - 1}{2 c_{12}}
      & - \frac{(c_{11} + 1)(c_{11} + 2)}{c_{12}}
      & - (c_{11} + 3)
    \end{pmatrix} ,
\end{equation}
and
\begin{equation}
  \setlength{\arraycolsep}{6pt}
  b^\text{p} =
    \begin{pmatrix}
      0 & 1 & 0 & 0 \\
      0 & 4 & 0 & 0 \\
      0 & 0 & 0 & 0 \\
      \frac{2 (k_{11} - 1)}{c_{12}}
      & - \frac{2 k_{11} + 4 k_{12} - 1}{2 c_{12}}
      & \frac{2 (c_{11} + 2)}{c_{12}}
      & 2
    \end{pmatrix} .
\end{equation}
If we choose the integration constants as
\begin{align}
  c_{12}=1, \qquad
  c_{11}=-1, \qquad
  k_{12}=\frac{1}{4}, \qquad
  k_{11}=-1 ,
\end{align}
we find agreement with the differential equation matrix given in
section~\ref{subsec:NNLOsol}, and the reconstructed $f_2^\text{p}$ turns
into $\text{V}_{12}$.

\subsubsection{Three or more coupled differential equations}
\label{subsubsec:3OrMore}

In the case that there are three or more coupled master integrals, one can
straightforwardly generalize the arguments in the above.
Suppose that one has $n$ coupled master integrals $(n \ge 2)$ to be added into
a canonical basis all at once.
Starting from the differential equations of order $m$ ($m \ge 1$)%
\footnote{%
 To keep the formulae simple, integrals that are already properly chosen as
 canonical master integrals and regarded as inhomogeneous terms are now also
 included in the basis vector $f$.
}%
\begin{equation}
  f^{(m)} = A^{[m-1]} f ,
\end{equation}
where the matrix $A^{[m]}$ is recursively defined by
\begin{equation}
  A^{[m]} := \left(A^{[m-1]}\right)' + A^{[m-1]} A ,
  \qquad
  A^{[0]} := A ,
\end{equation}
one can obtain the $n$th order differential equation for $f_i$ by eliminating
$(n-1)$ integrals $f_{j_1}$, $f_{j_2}$, $\dots$, $f_{j_{n-1}}$ from
the system of $n$ differential equations for
$f_i'$, $f_i^{(2)}$, \dots, $f_i^{(n)}$.
The result has the following form
\begin{equation}
  \sum_{m=1}^n C_m f_i^{(m)}
  = \sum_{k\notin\{j_1,\dots,j_{n-1}\}} C_{0k} f_k ,
\end{equation}
where the summation in the right-hand side is taken for all integrals
except the eliminated integrals;
in other words, all canonical master integrals already fixed as well as $f_i$.
The coefficients $C_m$ and $C_{0k}$ are given by
\begin{equation}
  C_m = \frac{\Delta_{mn}}{\Delta_{nn}} , \qquad
  C_{0k} = \frac{\Delta_k}{\Delta_{nn}}
    = \sum_{m=1}^n C_m A^{[m-1]}_{ik} ,
  \label{CmC0k}
\end{equation}
where we have normalized $C_n$ as unity,
$\Delta_k = \det(M_k)$ is the determinant of the following $n \times n$ matrix:
\begin{equation}
  M_k = \begin{pmatrix}
          A_{ij_1}^{[0]} & \hdots & A_{ij_{n-1}}^{[0]} & A_{ik}^{[0]} \\
        \vdots & & \vdots & \vdots\\
          A_{ij_1}^{[n-1]} & \hdots & A_{ij_{n-1}}^{[n-1]} & A_{ik}^{[n-1]}
        \end{pmatrix} ,
\end{equation}
and $\Delta_{mn}$ is the cofactor obtained by multiplying $(-1)^{m+n}$
to the determinant of $M_k$ with omitting the $m$th row and the $n$th column
(hence does not depend on $k$).

Assuming $f_i$ is a canonical master integral, one can conclude that
the coefficients must have the following structure in $\epsilon$:
\begin{align}
  C_m &= \frac{
           \epsilon^{n-1} C_m^{(n-1)} + \dots +
           \epsilon^{n(n+1)/2-m} C_m^{(n(n+1)/2-m)}
         }{
           \epsilon^{n-1} D^{(n-1)} + \dots +
           \epsilon^{n(n-1)/2} D^{(n(n-1)/2)}
         } , \nonumber \\
  C_{0k} &= \frac{
              \epsilon^{n} C_{0k}^{(n)} + \dots +
              \epsilon^{n(n+1)/2} C_{0k}^{(n(n+1)/2)}
            }{
              \epsilon^{n-1} D^{(n-1)} + \dots +
              \epsilon^{n(n-1)/2} D^{(n(n-1)/2)}
            } .
  \label{CmC0k-decomposition}
\end{align}
The coefficients $C_m$ and $C_{0k}$ are rational functions
in $\epsilon$ and thus the set of differential equations for reconstructing
the matrix $A^\text{p}$,
appearing in the basis in which the eliminated integrals $f_{j_1}$, $f_{j_2}$,
$\dots$, $f_{j_{n-1}}$ are properly chosen,
becomes quite tedious.
However, taking only the leading terms of $C_m$ and $C_{0k}$
in $\epsilon$ by replacing $A^{[n]}$ with $A^{(n)}$
\begin{equation}
  A^{[n]}_{kl} = A^{(n)}_{kl} + \mathcal{O}(\epsilon^2) ,
\end{equation}
may alleviate the complexity of the problem%
\footnote{%
  This does not apply to the cases where $C_m^{(n-1)}$, $D^{(n-1)}$ or
  $C_{0k}^{(n)}$ obtained from the components of $A^\text{p}$ become zero.
  For example,
  with the ansatz eq.~\eqref{aijHennmasters}, one finds
  $C_m^{(n-1)}$ and $D^{(n-1)}$ vanish for $n \ge 5$,
  and $C_{0k}^{(n)}$ vanishes for $n \ge 4$.
}%
.
The first terms of $C_m$, i.e., $C_m^{(n-1)}/D^{(n-1)}$ give a set of $(n-1)$ differential
equations of $(n-1)$ variables $A_{ij_1}^\text{p}$, $\dots$,
$A_{ij_{n-1}}^\text{p}$.
After solving them, one substitutes the result into the first terms of
$C_{0k}$, $C_{0k}^{(n)}/D^{(n-1)}$, which gives a differential equation for $A_{ik}^\text{p}$.
In this way, one can reconstruct the $i$th row of the matrix $A^\text{p}$.

In general, higher order terms of the coefficients in $\epsilon$ expansions are
needed to reconstruct the full matrix $A^\text{p}$.
From a naive counting, $n$ orders of each coefficient have to be taken
into account.

Once $A^\text{p}$ is completely determined,
one can construct the basis transformation $B$ to this basis.
The matrix $B$ can be parametrized by filling $(n-1)$ rows corresponding to
$f_{j_1}$, $\dots$, $f_{j_{n-1}}$ with variables $b_{kl}$.
The matrix equation eq.~\eqref{deqForBS} contains derivatives of the variables;
however, one does not need to solve any differential equations.
Non-trivial equations in $n$ rows of the matrix equation can be
solved as follows.
Starting with $i$th row, whose components are all zero in the left-hand side,
one has a set of linear equations,
which can be solved for all variables of a row%
\footnote{%
  There exist cases where some components of $A^\text{p}$ are zero, and
  some variables do not appear in a set of equations generated from a row
  of the matrix equation.
  However, the set of equations must give solutions for variables of a row
  at least provided the integral corresponding to the row giving the equations
  is coupled to the other integrals in the basis with $A^\text{p}$.
}%
.
Next, one considers this row.
On the left-hand side, one can use the chain rule of the derivative and
replace derivatives of unsolved variables with
the corresponding components of the matrix equation.
Then one substitutes the solution for the solved variables.
The resulting equations give
the next set of linear equations that can determine
all variables of another row.
Repeating this procedure,
one can solve for all the variables by using $(n-1)$ rows,
and the remaining row can serve as a consistency check.

The generalized version of the algorithm given in section~%
\ref{subsubsec:B-construction} that checks whether $f_i$ is a canonical master
integral is formulated as follows:
\begin{enumerate}
  \item Derive higher order differential equation for $f_i$,
        i.e. calculate the coefficients $C_m$ and $C_{0k}$
        given in eq.~\eqref{CmC0k}.
  \item Check $\epsilon$-dependence of $C_m$ and $C_{0k}$, which should be
        as in eq.~\eqref{CmC0k-decomposition}.
  \item Expand $C_m$ and $C_{0k}$ in $\epsilon$.
        Take enough terms to be able to solve for the elements of $A^\text{p}$.
        The differential equations should be solvable by the ansatz
        eq.~\eqref{aijHennmasters}.
        In order to proceed,
        it is enough to find one particular solution for $A^\text{p}$.
  \item Construct $B$ and check its consistency by use of eq.~\eqref{deqForBS}.
\end{enumerate}
If the checks fail at any step, one can conclude $f_i$ cannot be a canonical
master integral.

Let us conclude with a few final remarks:
\begin{itemize}
  \item By using the ansatz eq.~\eqref{aijHennmasters}
        in solving for the elements of $A^\text{p}$,
        all differential equations appearing in this algorithm can be
        reduced to algebraic equations.
  \item In practice $A$ sometimes contains elements equal to zero.
        Setting the elements at the same positions in $A^\text{p}$ to zero
        may simplify the derivation a lot, provided this additional
        constraints on the form of $A^\text{p}$ gives a solution for
        $A^\text{p}$ and $B$.
  \item By changing the ansatz eq.~\eqref{aijHennmasters},
        the algorithm can be extended to
        other forms~\eqref{AbarStruct} in $x$.
\end{itemize}

\subsection{Canonical master integrals for NNLO Higgs boson production}
\label{subsec:NNLOsol}

\begin{figure}[tbp]
  \centering
  \setlength{\tabcolsep}{6pt}
  \renewcommand{\arraystretch}{6}
  \renewcommand{\GraphSize}{.3}
  \def\ParseGraph#1#2#3#4-{%
    \def\ParseGraphh##1##2##3##4##5##6##7:{%
      $\text{#1#2#3}_{#4}(##1,##2,##3,##4,##5,##6,##7)$
    }%
    \ParseGraphh
  }
  \begin{tabular}{ccc}
    \IncGraph{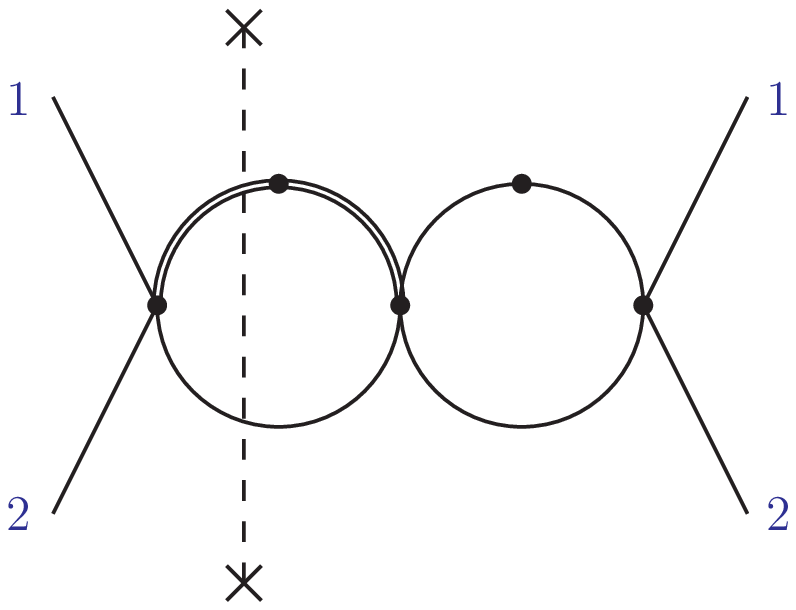} &
    \IncGraph{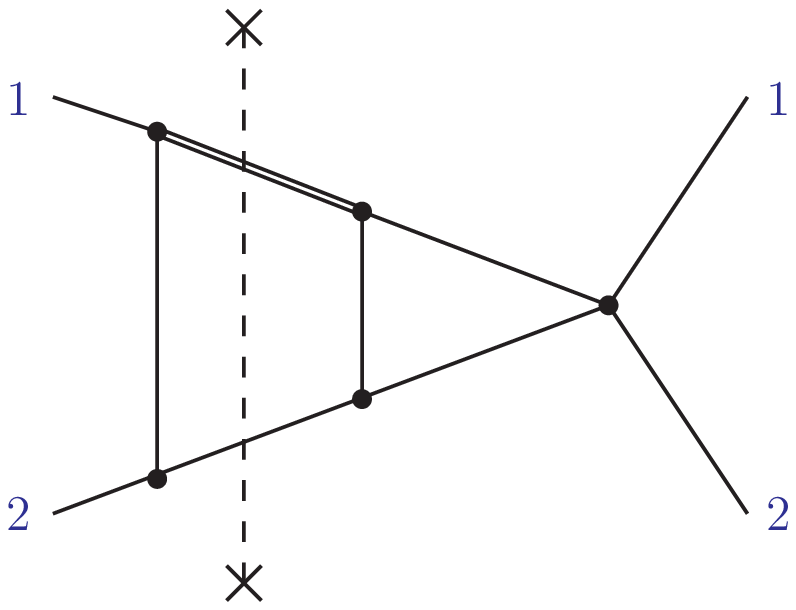} &
    \IncGraph{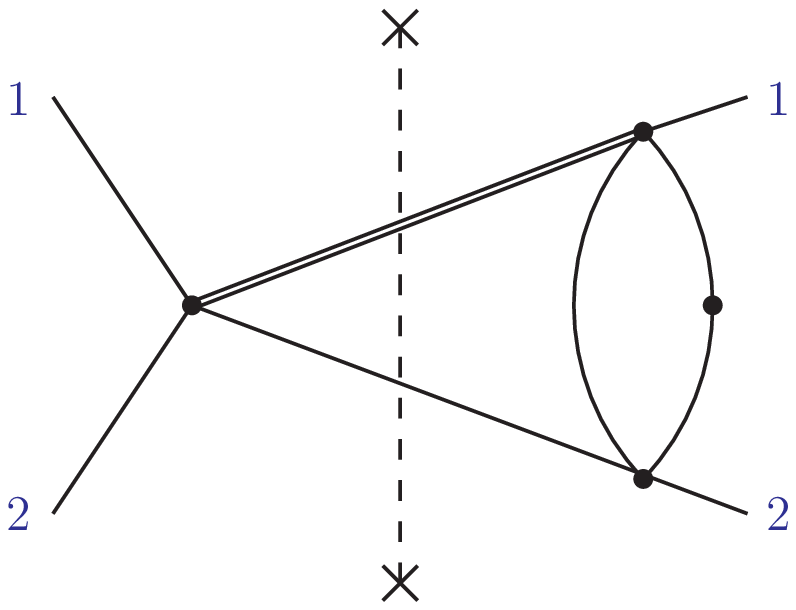} \\
    \IncGraph{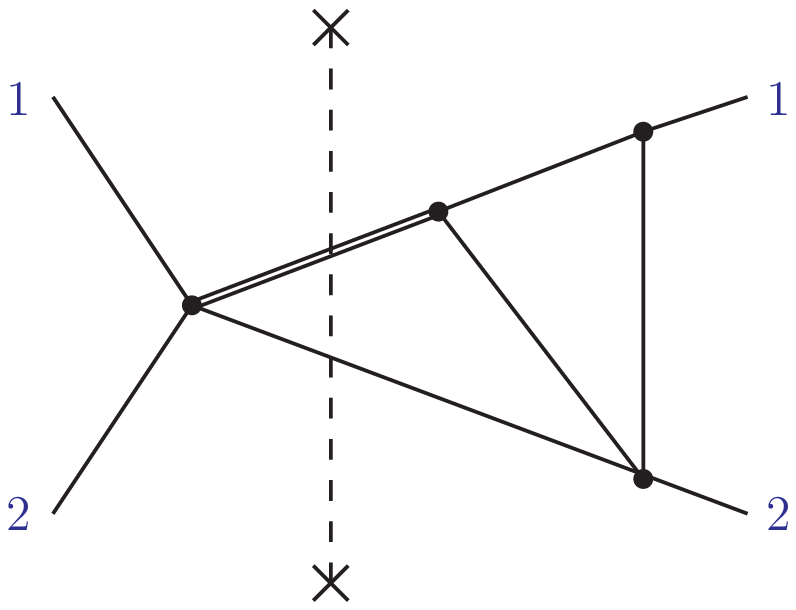} &
    \IncGraph{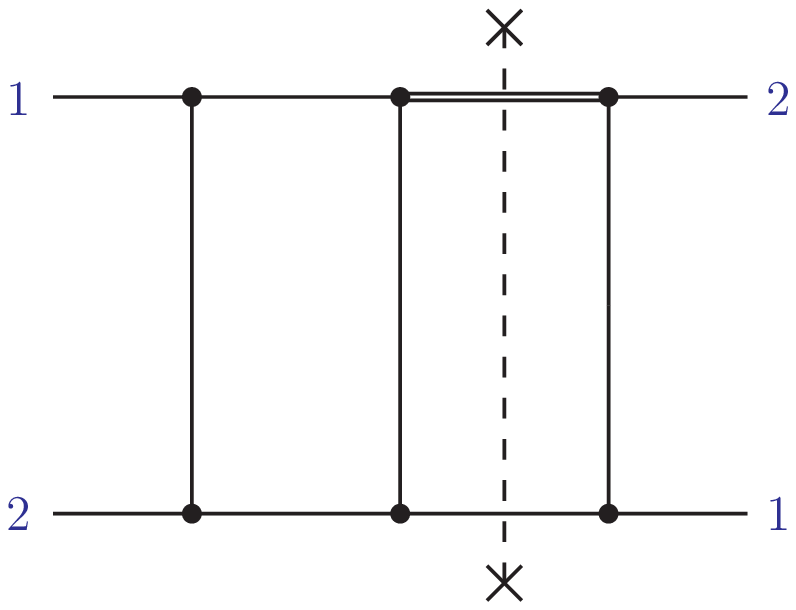} &
    \IncGraph{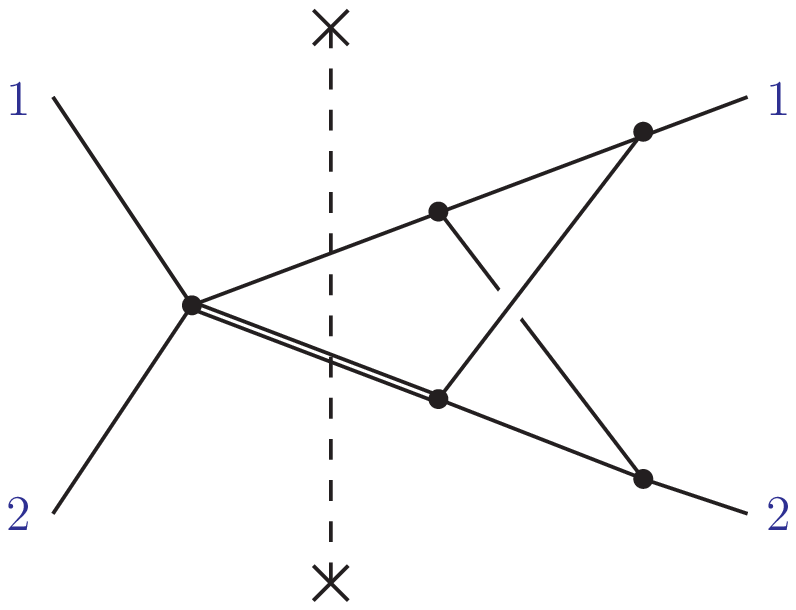}
  \end{tabular}
  \caption{
    \label{fig:NNLOdiags2}
    Two-particle cut diagrams appearing in our choice of
    canonical master integrals at NNLO.
  }
\end{figure}

\begin{figure}[tbp]
  \centering
  \setlength{\tabcolsep}{6pt}
  \renewcommand{\arraystretch}{1.5}
  \renewcommand{\GraphSize}{.3}
  \def\ParseGraph#1#2#3#4-{%
    \def\ParseGraphh##1##2##3##4##5##6##7:{%
      \vspace{-5pt}%
      $\text{#1#2#3}_{#4}(##1,##2,##3,##4,##5,##6,##7)$
    }%
    \ParseGraphh
  }
  \begin{tabular}{ccc}
    \IncGraph{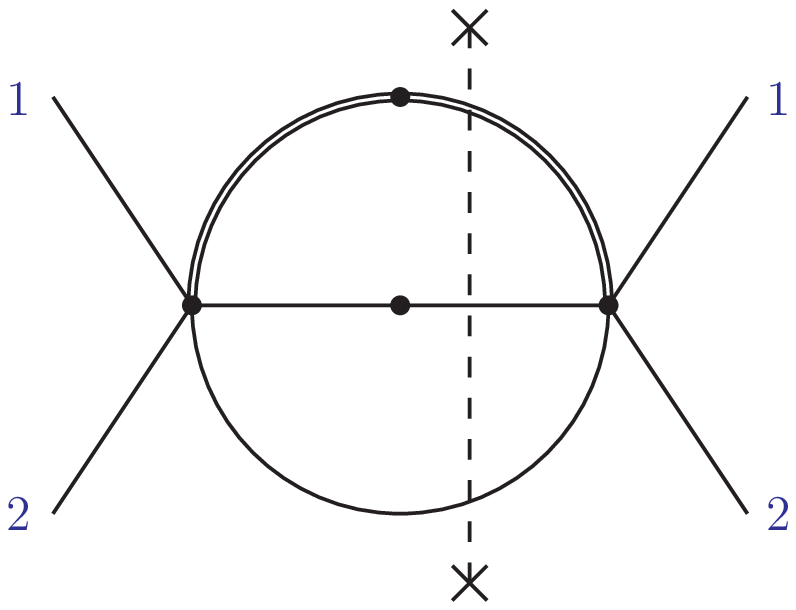} &
    \IncGraph{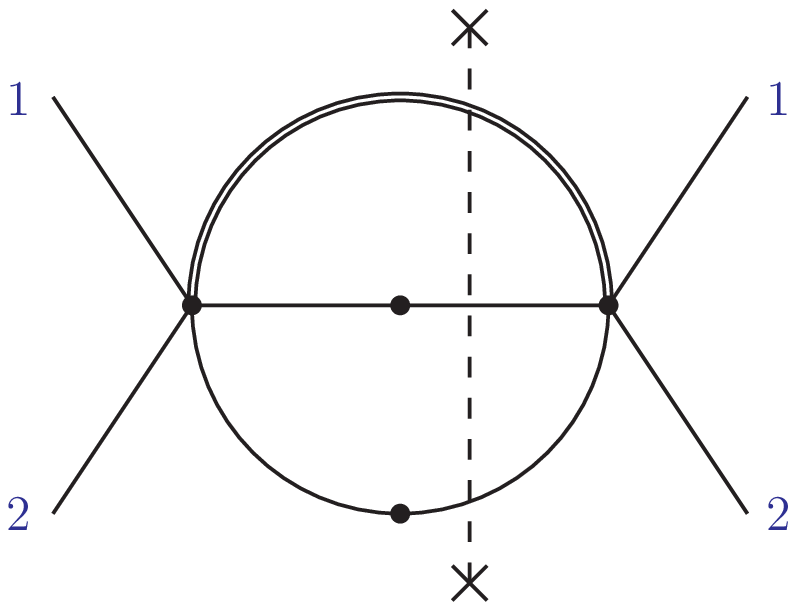} &
    \IncGraph{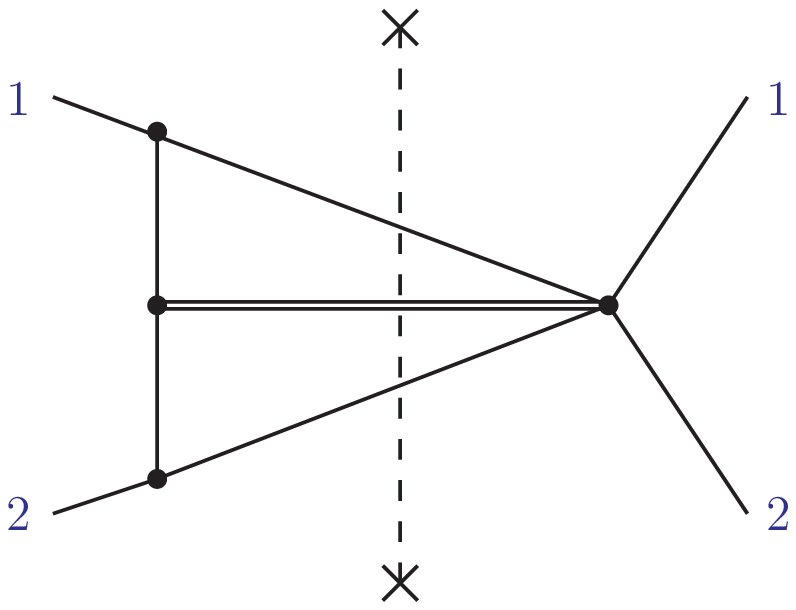} \\
    \IncGraph{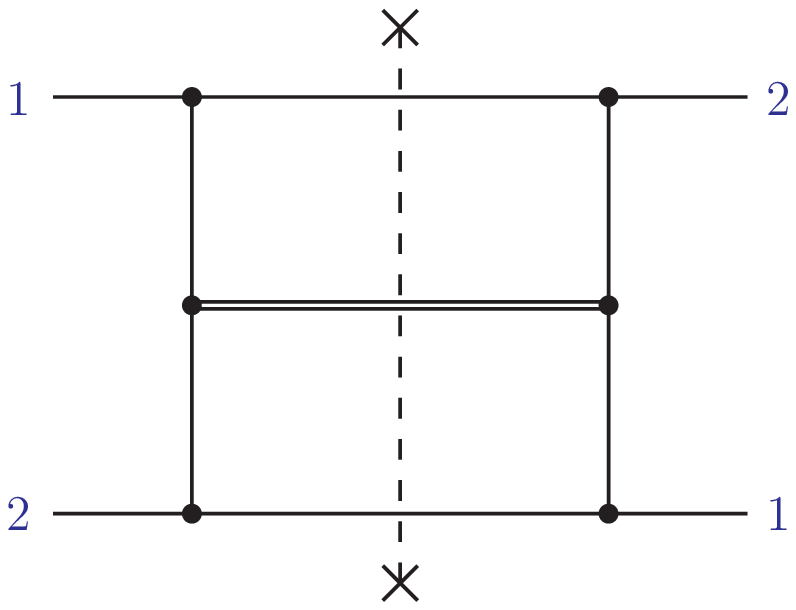} &
    \IncGraph{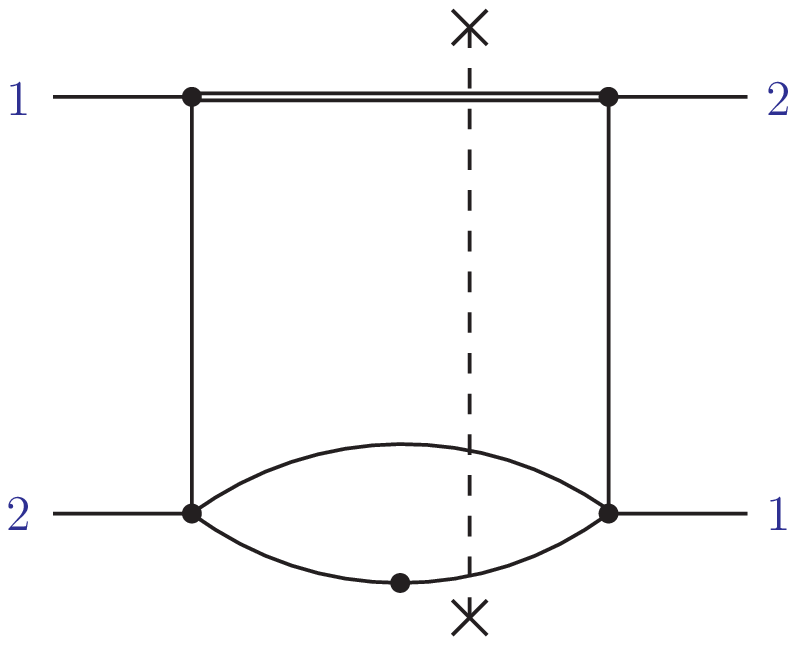} &
    \IncGraph{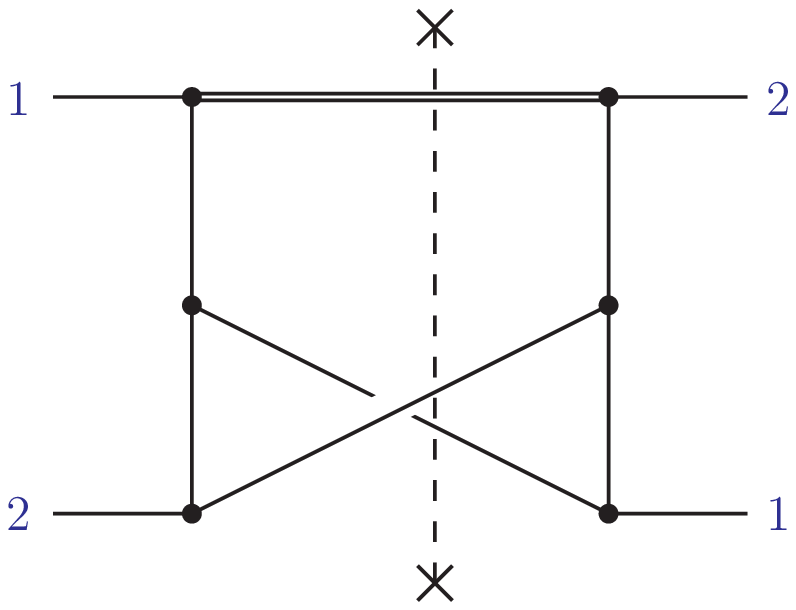} \\
    \IncGraph{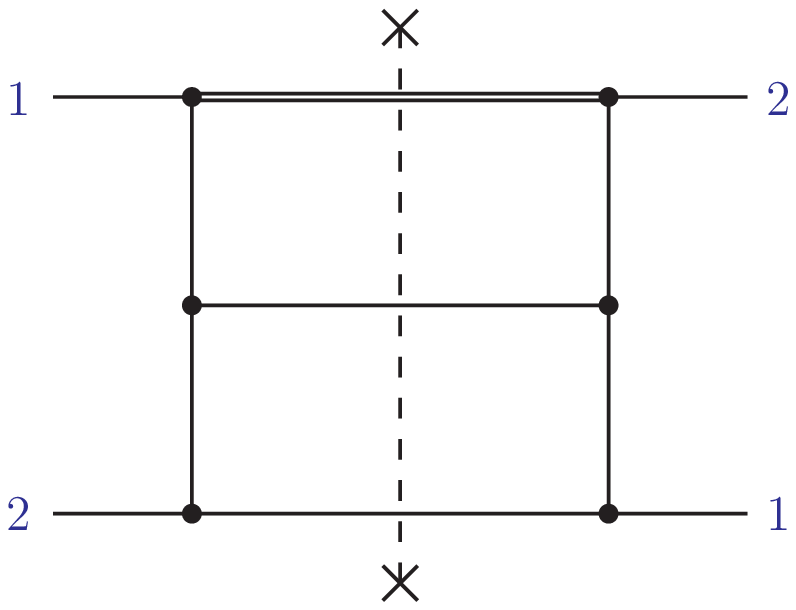} &
    \IncGraph{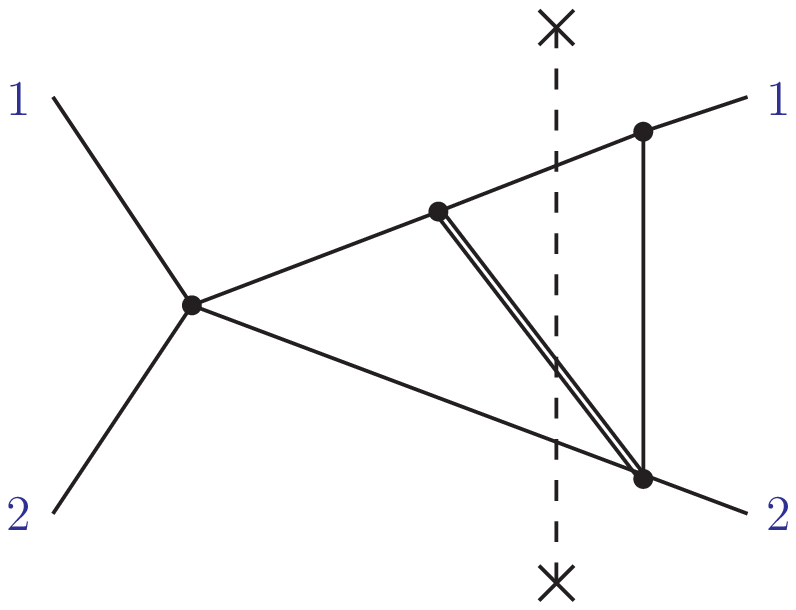} &
    \IncGraph{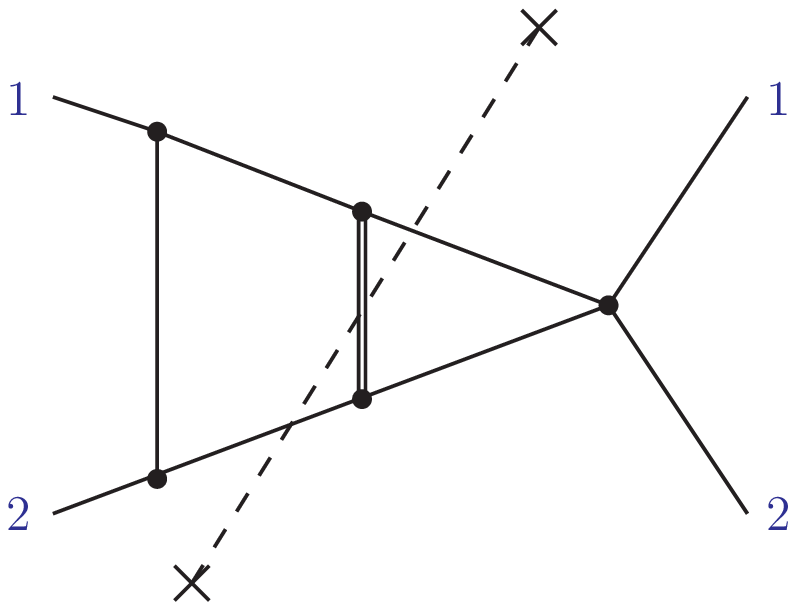} \\
    \IncGraph{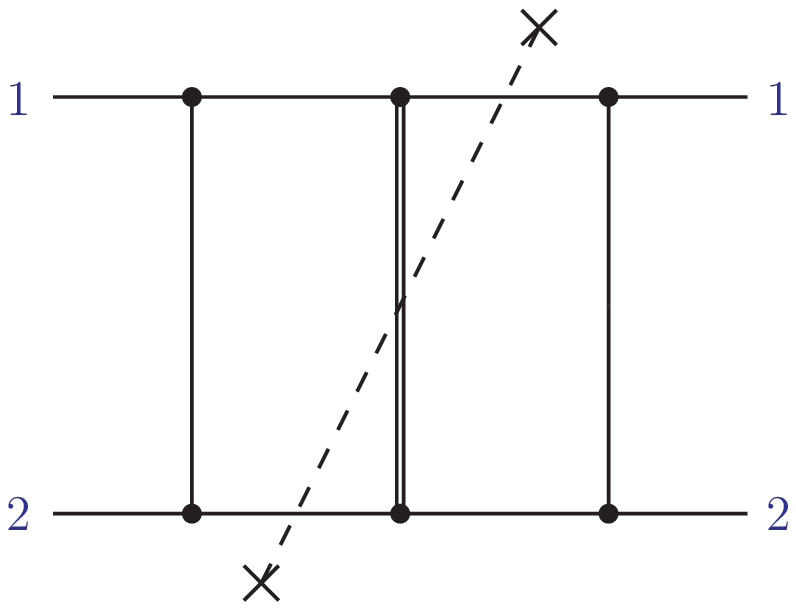} &
    \IncGraph{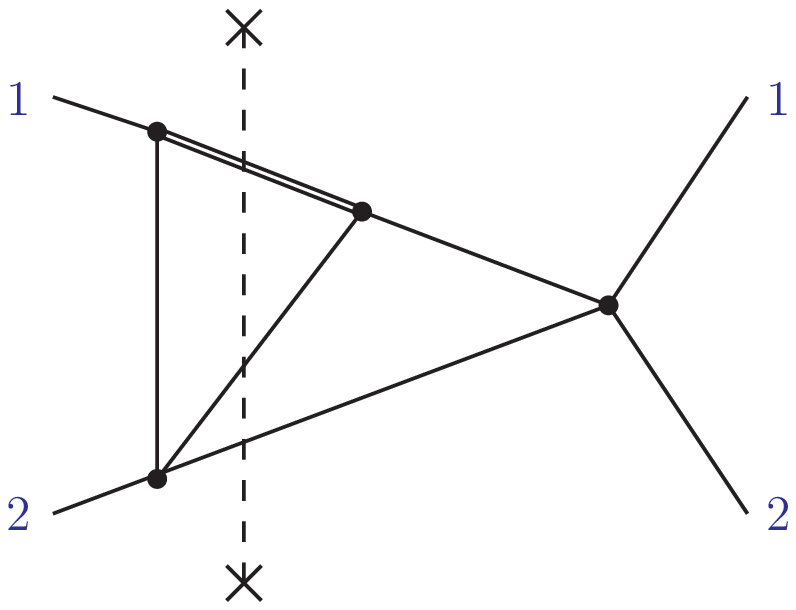} &
    \IncGraph{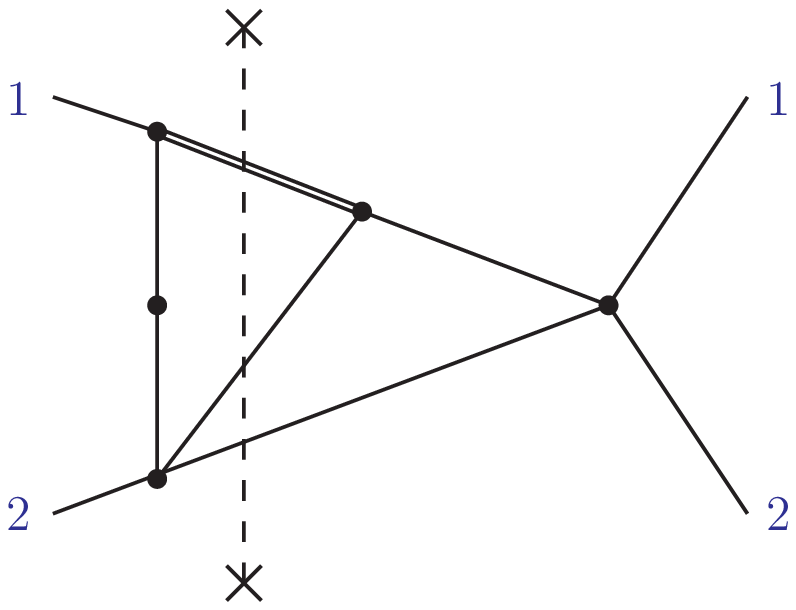} \\
    \IncGraph{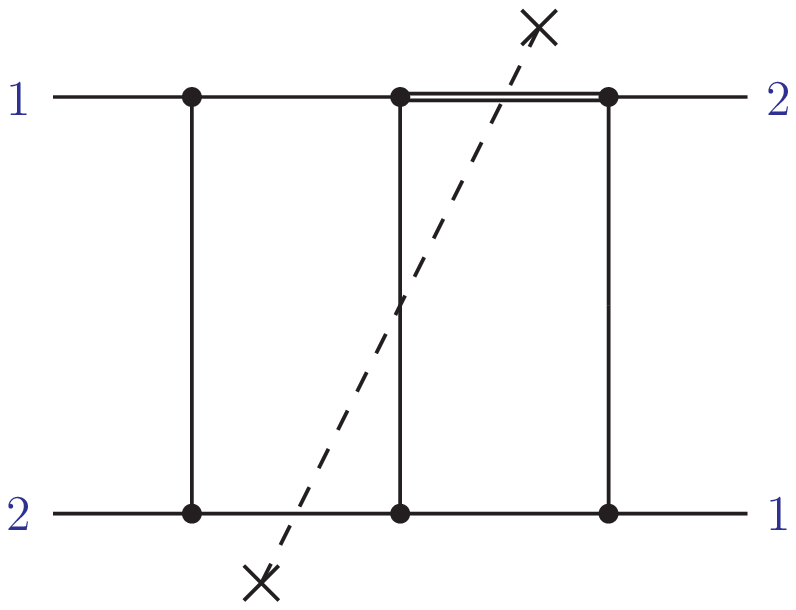} &
    \IncGraph{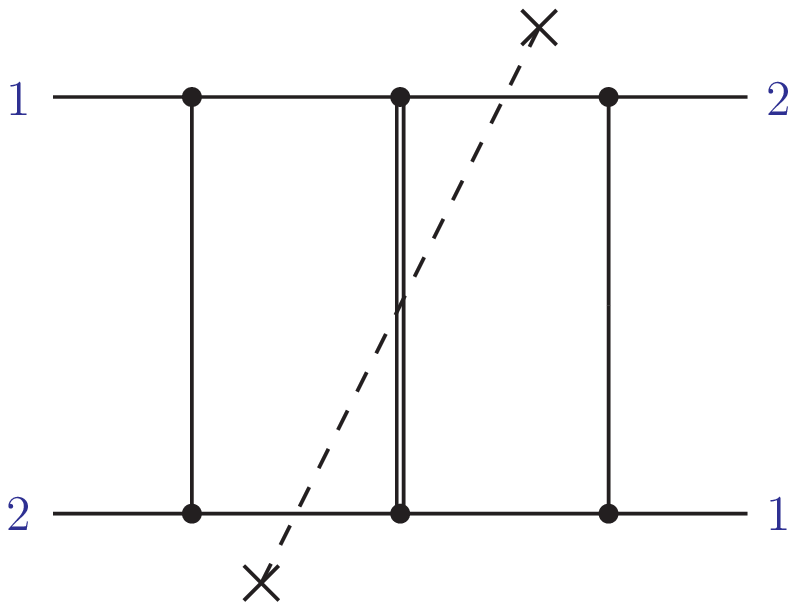} &
    \IncGraph{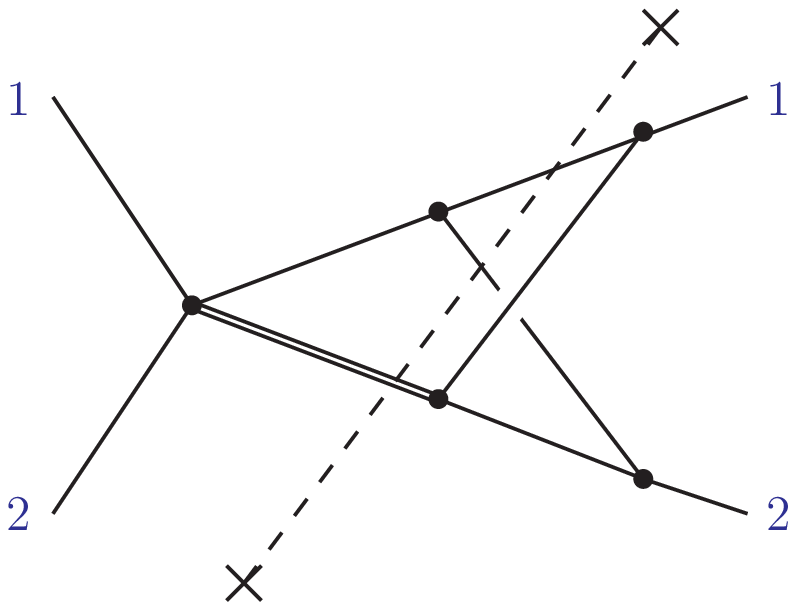} \\
    \IncGraph{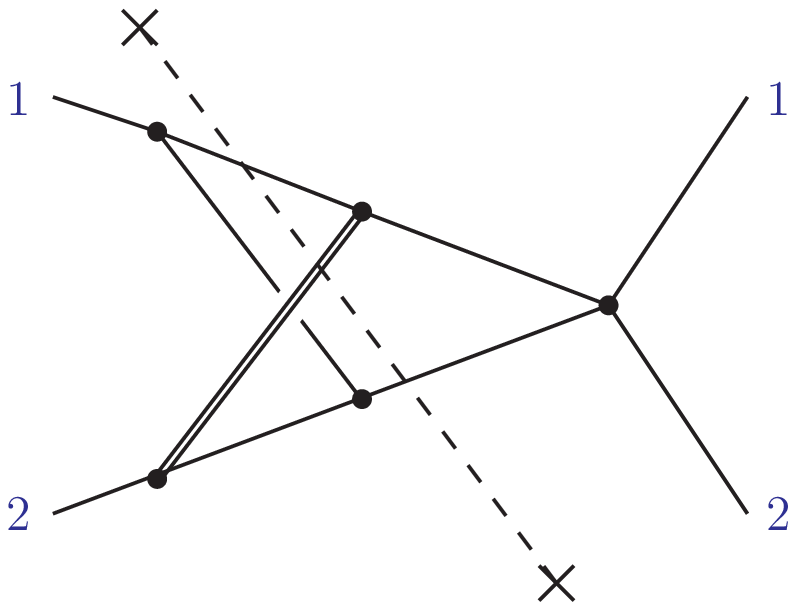} &
    \IncGraph{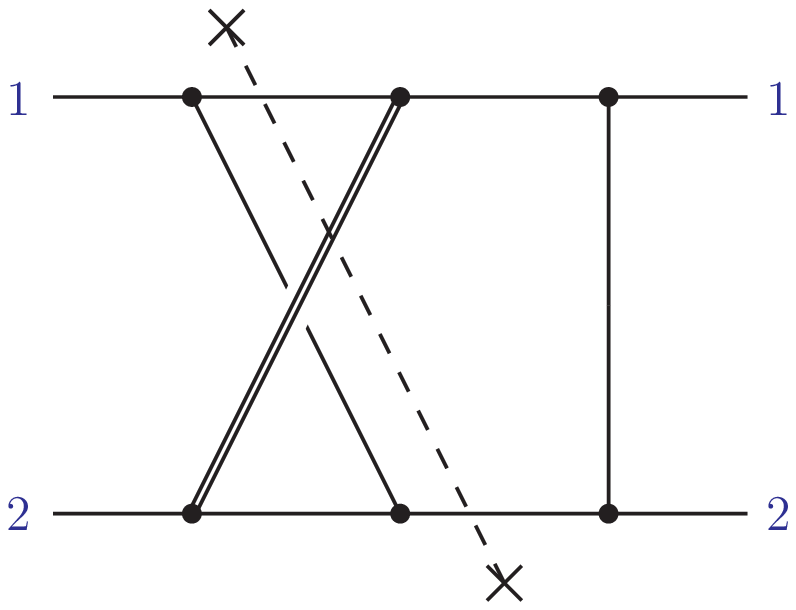}
  \end{tabular}
  \caption{
    \label{fig:NNLOdiags3}
    Three-particle cut diagrams appearing in our choice of
    canonical master integrals at NNLO.
  }
\end{figure}

Here we present a canonical basis we found at NNLO, together with
the differential equation matrix $A$ it satisfies.
All master integrals of topology $T$ in this basis have the form
\begin{align}\label{HPLmasterForm}
M^{(T)}_i = \epsilon^{d_i} n_i(x) \sum_k c_{ik} T\left(\{a\}_k\right),
\end{align}
where $d_i$ is an integer,
$n_i(x)$ is an $x$-dependent prefactor,
$c_{ik}$ are numerical constants
and $\{a\}_k$ are distinct sets of indices.
All integrals are defined as single-cut integrals.
The definitions of the individual topologies are given
in fig.~\ref{fig:NNLOtops}.
Note that the choice of a canonical basis is not unique.
In many cases we have found alternative master integrals forming a canonical basis
which have a more complicated $n(\epsilon)$ normalization.
We present here a basis of the simple monomial form in
$\epsilon$ as eq.~\eqref{HPLmasterForm}:
\begin{align}
 W_{1} &= \epsilon \text{TTF}_2(2,0,2,1,0,1,0), \nonumber\\
 W_{2} &= \epsilon^3 (1-x) \text{TTF}_2(1,0,1,1,1,1,1), \nonumber\\
 W_{3} &= \epsilon^2 \text{TTF}_2(1,2,0,0,1,1,0), \nonumber\\
 W_{4} &= \epsilon^3 \text{TTF}_2(1,1,1,0,1,1,0), \nonumber\\
 W_{5} &= \epsilon^3 (1-x) \text{TTG}_2(1,1,1,1,1,1,1), \nonumber\\
 W_{6} &= \epsilon^3 (1-x) \text{TTJ}_2(1,1,1,1,1,1,0),
\end{align}
are the two-particle cut master integrals, whereas
\begin{align}
 V_{1} &= \epsilon \text{TTA}_3(2,2,1,0,0,0,0), \nonumber\\
 V_{2} &= \epsilon (1-x) \bigl[\text{TTA}_3(1,2,2,0,0,0,0)+2 \text{TTA}_3(2,2,1,0,0,0,0)\bigr], \nonumber\\
 V_{3} &= \epsilon^3 \text{TTA}_3(1,1,1,1,0,1,0), \nonumber\\
 V_{4} &= \epsilon^3 (1-x) \text{TTA}_3(1,1,1,1,1,1,1), \nonumber\\
 V_{5} &= \epsilon^2 (1-x) \text{TTC}_3(1,2,1,0,1,1,0), \nonumber\\
 V_{6} &= \epsilon^3 (1-x) x \text{TTC}_3(1,1,1,1,1,1,1), \nonumber\\
 V_{7} &= \epsilon^3 (1-x) \text{TTD}_3(1,1,1,1,1,1,1), \nonumber\\
 V_{8} &= \epsilon^3 \text{TTE}_3(1,0,1,1,1,0,1), \nonumber\\
 V_{9} &=
\epsilon^3 \text{TTE}_3(1,1,1,1,1,1,0), \nonumber\\
 V_{10} &=
\epsilon^3 x \text{TTE}_3(1,1,1,1,1,1,1), \nonumber\\
 V_{11} &= \epsilon^3 \text{TTF}_3(1,0,1,1,1,0,1), \nonumber\\
 V_{12} &= \epsilon^2 x \text{TTF}_3(1,0,1,1,1,0,2), \nonumber\\
 V_{13} &= \epsilon^3 (1-x) \text{TTG}_3(1,1,1,1,1,1,1), \nonumber\\
 V_{14} &=
\epsilon^3 (1+x) \text{TTH}_3(1,1,1,1,1,1,1), \nonumber\\
 V_{15} &= \epsilon^3 (1-x) \text{TTJ}_3(1,1,1,1,1,1,0), \nonumber\\
 V_{16} &= \epsilon^3 (1+x) \text{TTK}_3(1,1,1,1,1,1,0), \nonumber\\
 V_{17} &= \epsilon^3 x \bigl[\text{TTK}_3(1,1,1,1,1,1,1)-\text{TTK}_3(1,1,1,1,1,1,0)\bigr],
\end{align}
are the three-particle cut master integrals\footnote{%
The basis given here has the
same number of integrals as the reduction basis given in ref.~\cite{Pak:2011hs} which is
known to be not minimal as there is a linear relation between the integrals
$U_{1}, U_{1\text{a}}, U_6$ and $U_8$ given there.}.
They are normalized in such a way as to start at finite order.
Two- and three-particle cut diagrams appearing in this basis are
shown in figs.~\ref{fig:NNLOdiags2} and~\ref{fig:NNLOdiags3}, respectively.

The two-particle cut master integrals satisfy the differential equations%
~\eqref{HennDEQ} and~\eqref{AForm} with
\begin{align}
a_2=
\begin{pmatrix}
 0 & 0 & 0 & 0 & 0 & 0 \\
 1 & -1 & 0 & 0 & 0 & 0 \\
 0 & 0 & 0 & 0 & 0 & 0 \\
 1 & -1 & -2 & -2 & 0 & 0 \\
 2 & -2 & 2 & 2 & 0 & 0 \\
 0 & 0 & -4 & -4 & 0 & 0 \\
\end{pmatrix},
\end{align}
\begin{align}
b_2=
\begin{pmatrix}
 2 & 0 & 0 & 0 & 0 & 0 \\
 0 & 2 & 0 & 0 & 0 & 0 \\
 0 & 0 & 3 & 0 & 0 & 0 \\
 0 & 0 & 0 & 0 & 0 & 0 \\
 0 & 0 & 6 & 0 & 2 & 0 \\
 0 & 0 & 0 & 0 & 0 & 4 \\
\end{pmatrix},
\end{align}
\begin{align}
c_2=0,
\end{align}
whereas for the three-particle cut master integrals we have
\begin{align}
a_3=
\begin{pmatrix}
 -3 & 1 & 0 & 0 & 0 & 0 & 0 & 0 & 0 & 0 & 0 & 0 & 0 & 0 & 0 & 0 & 0 \\
 -6 & 2 & 0 & 0 & 0 & 0 & 0 & 0 & 0 & 0 & 0 & 0 & 0 & 0 & 0 & 0 & 0 \\
 -1 & \frac{1}{2} & -2 & 0 & 0 & 0 & 0 & 0 & 0 & 0 & 0 & 0 & 0 & 0 & 0 & 0 & 0 \\
 -2 & 1 & -4 & 0 & 0 & 0 & 0 & 0 & 0 & 0 & 0 & 0 & 0 & 0 & 0 & 0 & 0 \\
 0 & -\frac{1}{2} & 0 & 0 & -1 & 0 & 0 & 0 & 0 & 0 & 0 & 0 & 0 & 0 & 0 & 0 & 0 \\
 -1 & \frac{1}{2} & -2 & 0 & 0 & -2 & 0 & 0 & 0 & 0 & 0 & 0 & 0 & 0 & 0 & 0 & 0 \\
 0 & 1 & 0 & 0 & 2 & 0 & 0 & 0 & 0 & 0 & 0 & 0 & 0 & 0 & 0 & 0 & 0 \\
 -2 & \frac{1}{2} & 0 & 0 & 0 & 0 & 0 & -2 & 0 & 0 & 0 & 0 & 0 & 0 & 0 & 0 & 0 \\
 4 & -1 & 0 & 0 & 0 & 0 & 0 & 1 & -1 & 0 & 0 & 0 & 0 & 0 & 0 & 0 & 0 \\
 -5 & \frac{3}{2} & 0 & 0 & 0 & 0 & 0 & -2 & 0 & -2 & 0 & 0 & 0 & 0 & 0 & 0 & 0 \\
 -1 & \frac{1}{4} & 0 & 0 & 0 & 0 & 0 & 0 & 0 & 0 & -1 & 1 & 0 & 0 & 0 & 0 & 0 \\
 -\frac{3}{2} & \frac{1}{2} & 0 & 0 & 0 & 0 & 0 & 0 & 0 & 0 & 0 & -2 & 0 & 0 & 0 & 0 & 0 \\
 -4 & \frac{3}{2} & 0 & 0 & 0 & 0 & 0 & 0 & 0 & 0 & 2 & 2 & 0 & 0 & 0 & 0 & 0 \\
 2 & 0 & 0 & 0 & 0 & 0 & 0 & 2 & -2 & 0 & 0 & 0 & 0 & 0 & 0 & 0 & 0 \\
 1 & -\frac{1}{2} & 2 & 0 & 0 & 0 & 0 & 0 & 0 & 0 & 0 & 0 & 0 & 0 & 0 & 0 & 0 \\
 3 & -1 & -2 & 0 & 0 & 0 & 0 & 0 & 0 & 0 & -2 & -2 & 0 & 0 & 0 & 0 & 0 \\
 1 & -\frac{1}{2} & 2 & 0 & 0 & 0 & 0 & 0 & 0 & 0 & 0 & 0 & 0 & 0 & 0 & 0 & -2 \\
\end{pmatrix},
\end{align}
\begin{align}
b_3=
\begin{pmatrix}
 0 & 1 & 0 & 0 & 0 & 0 & 0 & 0 & 0 & 0 & 0 & 0 & 0 & 0 & 0 & 0 & 0 \\
 0 & 4 & 0 & 0 & 0 & 0 & 0 & 0 & 0 & 0 & 0 & 0 & 0 & 0 & 0 & 0 & 0 \\
 0 & 0 & 0 & 0 & 0 & 0 & 0 & 0 & 0 & 0 & 0 & 0 & 0 & 0 & 0 & 0 & 0 \\
 0 & 2 & 0 & 2 & 0 & 0 & 0 & 0 & 0 & 0 & 0 & 0 & 0 & 0 & 0 & 0 & 0 \\
 0 & -\frac{1}{2} & 0 & 0 & 2 & 0 & 0 & 0 & 0 & 0 & 0 & 0 & 0 & 0 & 0 & 0 & 0 \\
 0 & 0 & 0 & 0 & -2 & 2 & 0 & 0 & 0 & 0 & 0 & 0 & 0 & 0 & 0 & 0 & 0 \\
 0 & 1 & 0 & 0 & 2 & 0 & 2 & 0 & 0 & 0 & 0 & 0 & 0 & 0 & 0 & 0 & 0 \\
 0 & 0 & 0 & 0 & 0 & 0 & 0 & 0 & 0 & 0 & 0 & 0 & 0 & 0 & 0 & 0 & 0 \\
 0 & 0 & 0 & 0 & 0 & 0 & 0 & 0 & 0 & 0 & 0 & 0 & 0 & 0 & 0 & 0 & 0 \\
 0 & 0 & 0 & 0 & 0 & 0 & 0 & 0 & 0 & 0 & 0 & 0 & 0 & 0 & 0 & 0 & 0 \\
 0 & 0 & 0 & 0 & 0 & 0 & 0 & 0 & 0 & 0 & 0 & 0 & 0 & 0 & 0 & 0 & 0 \\
 -4 & 1 & 0 & 0 & 0 & 0 & 0 & 0 & 0 & 0 & 2 & 2 & 0 & 0 & 0 & 0 & 0 \\
 -4 & \frac{3}{2} & 0 & 0 & 0 & 0 & 0 & 0 & 0 & 0 & 2 & 2 & 2 & 0 & 0 & 0 & 0 \\
 0 & 0 & 0 & 0 & 0 & 0 & 0 & 0 & 0 & 0 & 0 & 0 & 0 & 0 & 0 & 0 & 0 \\
 0 & 0 & 0 & 0 & 0 & 0 & 0 & 0 & 0 & 0 & 0 & 0 & 0 & 0 & 4 & 0 & 0 \\
 0 & 0 & 0 & 0 & 0 & 0 & 0 & 0 & 0 & 0 & 0 & 0 & 0 & 0 & 0 & 0 & 0 \\
 4 & -\frac{3}{2} & 0 & 0 & 0 & 0 & 0 & 0 & 0 & 0 & -2 & -2 & 0 & 0 & 0 & 0 & 0 \\
\end{pmatrix},
\end{align}
\begin{align}
c_3=
\begin{pmatrix}
 0 & 0 & 0 & 0 & 0 & 0 & 0 & 0 & 0 & 0 & 0 & 0 & 0 & 0 & 0 & 0 & 0 \\
 0 & 0 & 0 & 0 & 0 & 0 & 0 & 0 & 0 & 0 & 0 & 0 & 0 & 0 & 0 & 0 & 0 \\
 0 & 0 & 0 & 0 & 0 & 0 & 0 & 0 & 0 & 0 & 0 & 0 & 0 & 0 & 0 & 0 & 0 \\
 0 & 0 & 0 & 0 & 0 & 0 & 0 & 0 & 0 & 0 & 0 & 0 & 0 & 0 & 0 & 0 & 0 \\
 0 & 0 & 0 & 0 & 0 & 0 & 0 & 0 & 0 & 0 & 0 & 0 & 0 & 0 & 0 & 0 & 0 \\
 0 & 0 & 0 & 0 & 0 & 0 & 0 & 0 & 0 & 0 & 0 & 0 & 0 & 0 & 0 & 0 & 0 \\
 0 & 0 & 0 & 0 & 0 & 0 & 0 & 0 & 0 & 0 & 0 & 0 & 0 & 0 & 0 & 0 & 0 \\
 0 & 0 & 0 & 0 & 0 & 0 & 0 & 0 & 0 & 0 & 0 & 0 & 0 & 0 & 0 & 0 & 0 \\
 -4 & 1 & 0 & 0 & 0 & 0 & 0 & -1 & -1 & 0 & 0 & 0 & 0 & 0 & 0 & 0 & 0 \\
 8 & -2 & 0 & 0 & 0 & 0 & 0 & 2 & 2 & 0 & 0 & 0 & 0 & 0 & 0 & 0 & 0 \\
 0 & 0 & 0 & 0 & 0 & 0 & 0 & 0 & 0 & 0 & 0 & 0 & 0 & 0 & 0 & 0 & 0 \\
 0 & 0 & 0 & 0 & 0 & 0 & 0 & 0 & 0 & 0 & 0 & 0 & 0 & 0 & 0 & 0 & 0 \\
 0 & 0 & 0 & 0 & 0 & 0 & 0 & 0 & 0 & 0 & 0 & 0 & 0 & 0 & 0 & 0 & 0 \\
 0 & 0 & 0 & 0 & 0 & 0 & 0 & 0 & 0 & 0 & 0 & 0 & 0 & -2 & 0 & 0 & 0 \\
 0 & 0 & 0 & 0 & 0 & 0 & 0 & 0 & 0 & 0 & 0 & 0 & 0 & 0 & 0 & 0 & 0 \\
 0 & 0 & 0 & 0 & 0 & 0 & 0 & 0 & 0 & 0 & 0 & 0 & 0 & 0 & 0 & -4 & 0 \\
 0 & 0 & 0 & 0 & 0 & 0 & 0 & 0 & 0 & 0 & 0 & 0 & 0 & 0 & 0 & 2 & 0 \\
\end{pmatrix}.
\end{align}

As discussed previously, $(V_1, V_2)$ and $(V_{11}, V_{12})$
form coupled differential equations. The other integrals form a triangular system,
hence one can add a candidate integral to a subset of a canonical basis and
check if it successfully gives a larger canonical basis or not,
approaching a whole canonical basis step by step.
Some of the integrals in the canonical basis given here are diagrammatically
similar to those given in
ref.~\cite{Henn:2013pwa} where four-point two-loop diagrams have been investigated
as well, although with different kinematics.

We computed the given master integrals
$W_i$ and $V_i$ up to order $\epsilon^6$ corresponding to a maximum weight of six
in the appearing HPLs,
using their reduction to the reduction basis and
the $x\to 1$ limit of the latter as boundary conditions.
We checked that the solutions for the master integrals in the canonical basis are
pure functions.
Furthermore, by applying the matrix $B^{-1}$ we transformed back to
the reduction basis and found agreement with the results given in
ref.~\cite{Pak:2011hs} up to the order in $\epsilon$ given there
which is high enough for the \NNNLO{} calculation
(see also the result in ref.~\cite{Anastasiou:2012kq}).
We emphasize that in the canonical basis master integrals decouple order by order in
$\epsilon$ and therefore only the first term in the $x\to 1$ limit
for each master integral is sufficient to fix all the boundary constants.

\section{\texorpdfstring{\NNNLO{}}{NNNLO}: example and solution}
\label{sec:NNNLO}

\begin{figure}[tbp]
  \centering
  \includegraphics[width=.4\textwidth,clip]{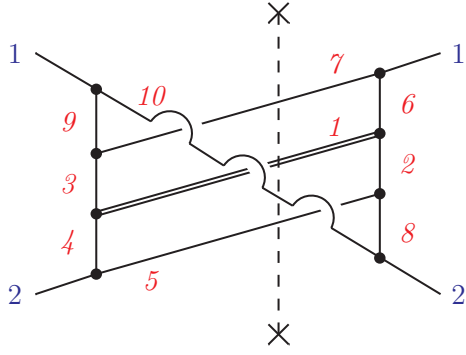}
  \hfill
  \caption{
    \label{fig:SeaSnake}
    The sea snake topology
    $\text{TTS}_4(a_1,a_2,a_3,a_4,a_5,a_6,a_7,a_8,a_9,a_{10},n_{11},n_{12})$.
    The notation is the same as in fig.~\ref{fig:NNLOtops}.
    The indices $n_{11}$ and $n_{12}$ denote irreducible scalar products
    which appear in the numerator and are always less than or equal to zero.
  }
\end{figure}

\begin{figure}[tbp]
  \centering
  \setlength{\tabcolsep}{6pt}
  \renewcommand{\arraystretch}{6}
  \renewcommand{\GraphSize}{.3}
  \def\ParseGraph#1#2#3#4-{%
    \def\ParseGraphh##1##2##3##4##5##6##7##8##9{%
      \def\ParseGraphhh####1####2####3:{%
        $\text{#1#2#3}_{#4}$%
        \scalebox{0.95}{(##1,##2,##3,##4,##5,##6,##7,##8,##9,####1,####2,####3)}%
      }%
      \ParseGraphhh
    }%
    \ParseGraphh
  }
  \begin{tabular}{ccc}
    \IncGraph{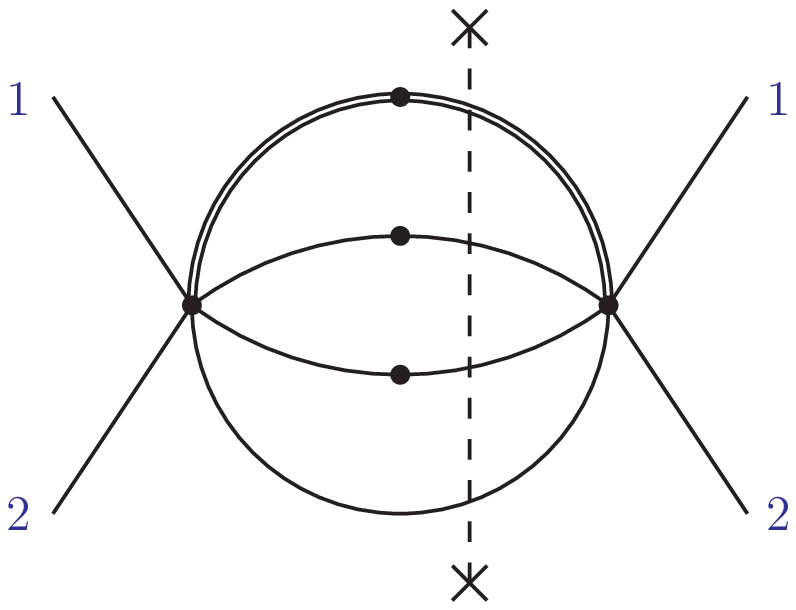} &
    \IncGraph{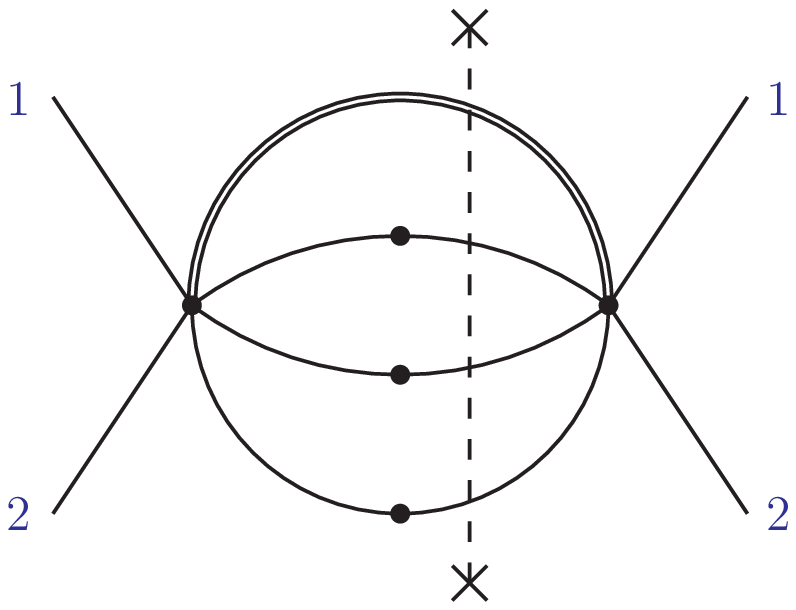} &
    \IncGraph{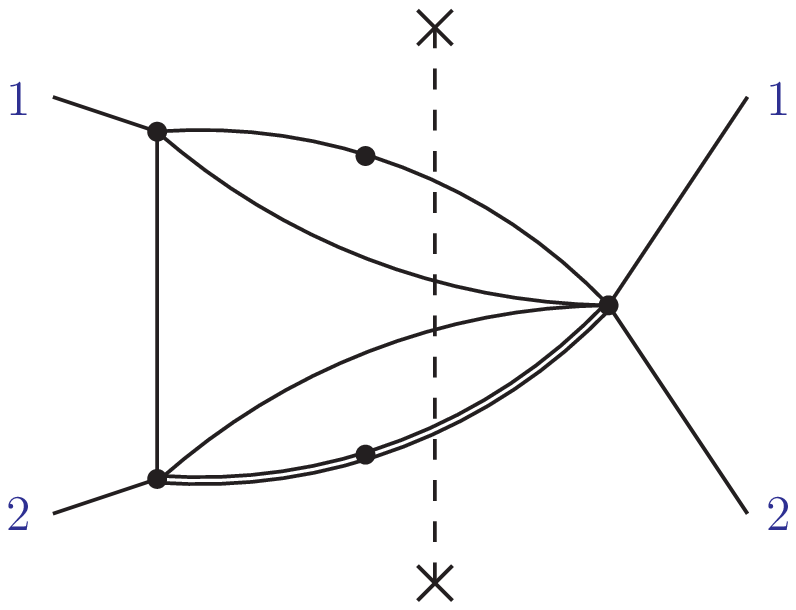} \\
    \IncGraph{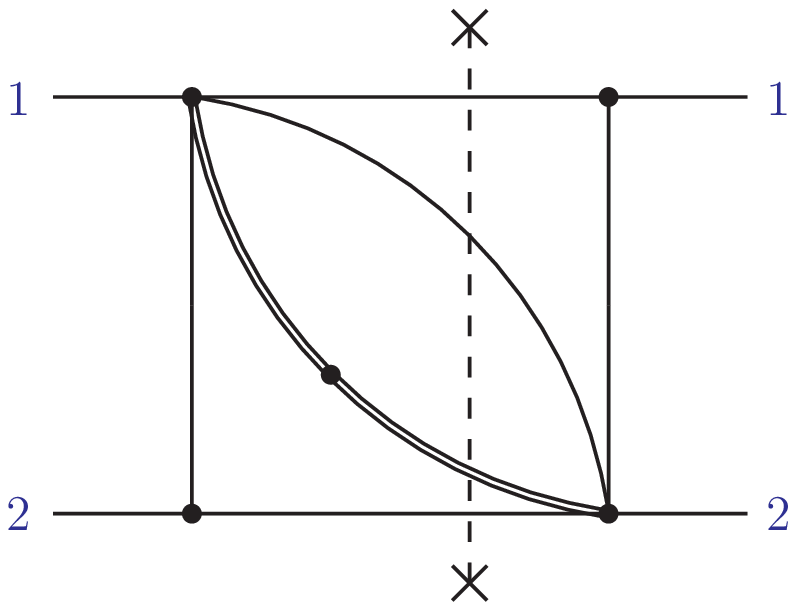} &
    \IncGraph{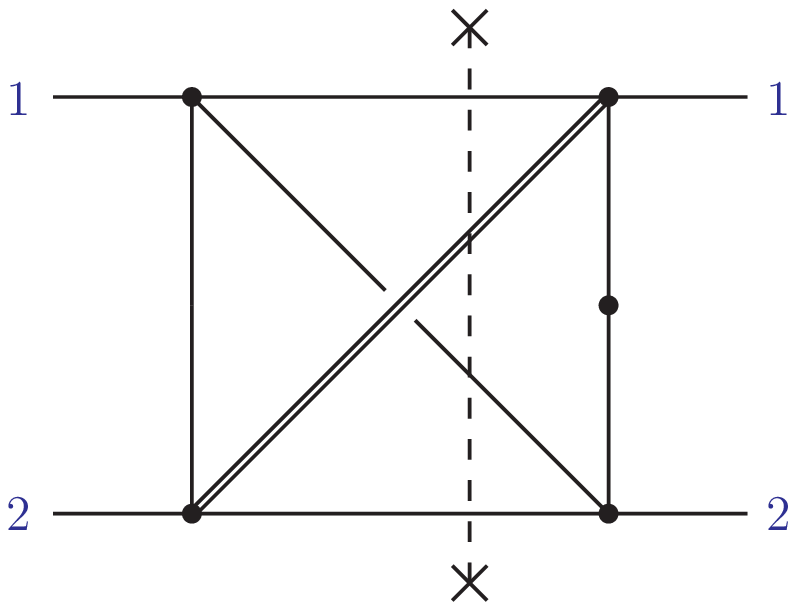} &
    \IncGraph{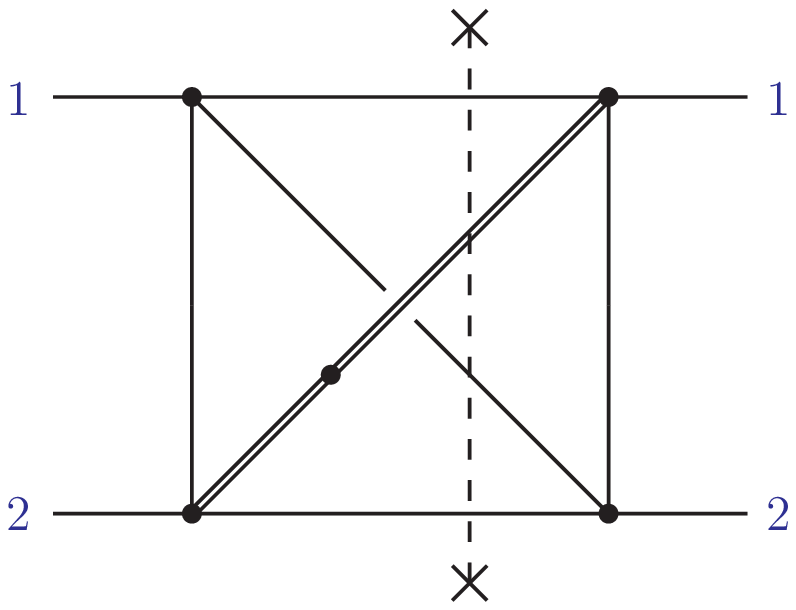} \\
    \IncGraph{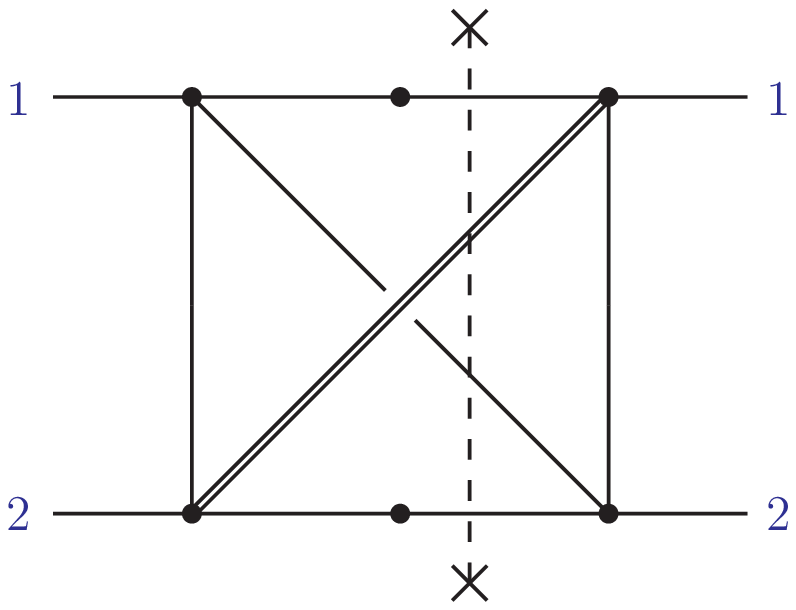} &
    \IncGraph{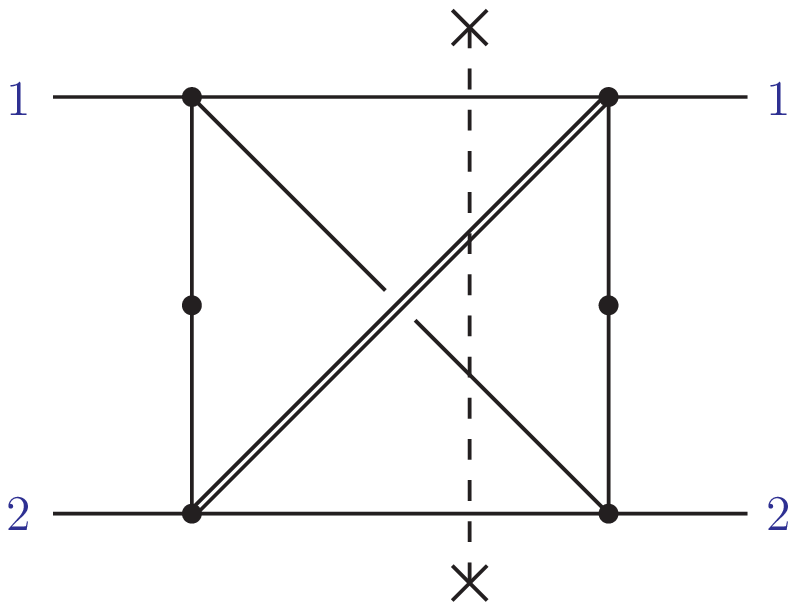} &
    \IncGraph{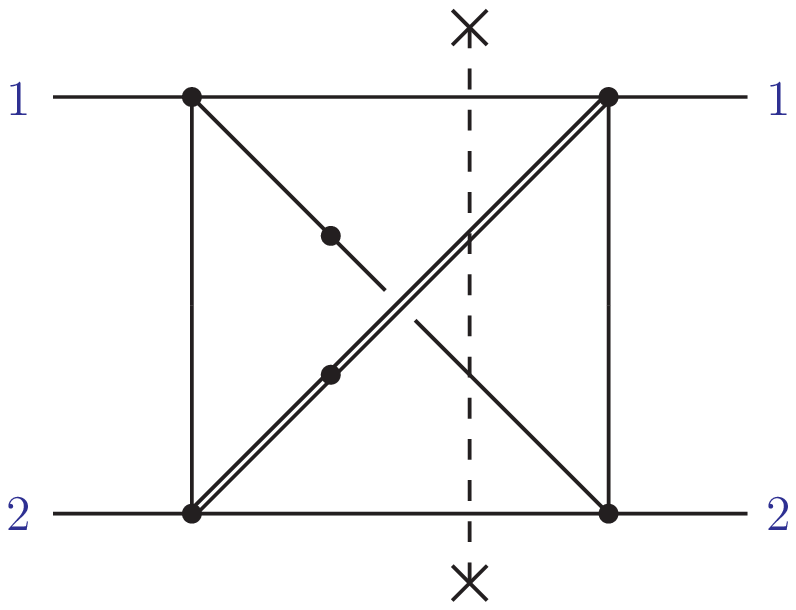} \\
    \IncGraph{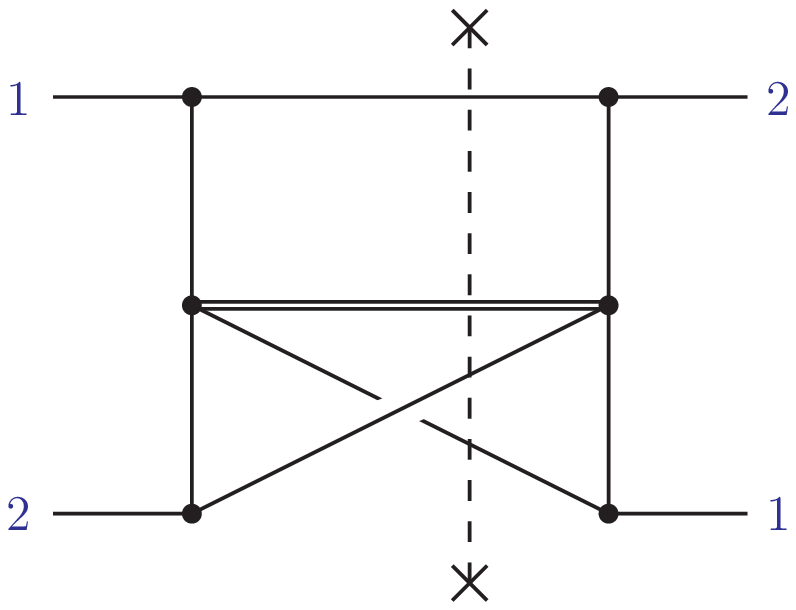} &
    \IncGraph{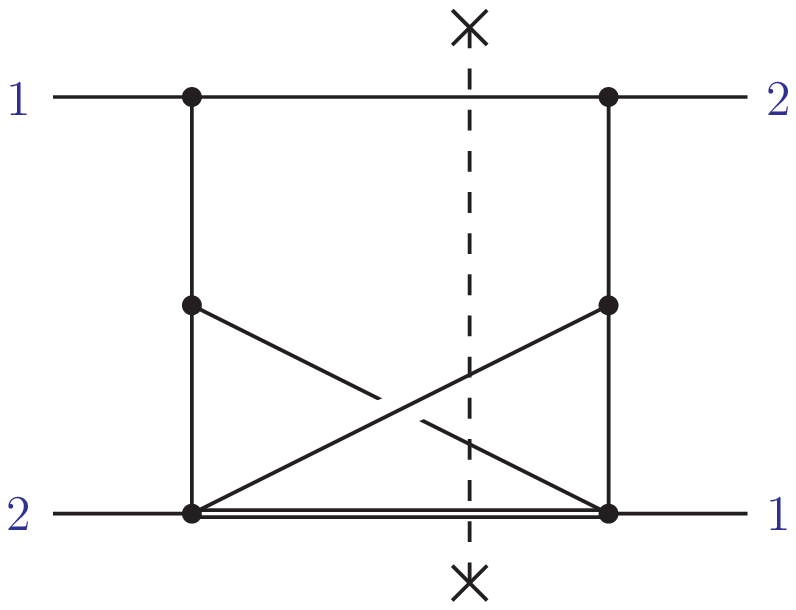}
  \end{tabular}
  \caption{
    \label{fig:NNNLOdiags}
    Diagrams appearing in our choice of canonical master integrals
    for the sea snake topology.
  }
\end{figure}

In this section we discuss the \NNNLO{} topology $\text{TTS}_4$, shown in fig.~\ref{fig:SeaSnake},
which we refer to as \textit{sea snake} topology.
It is non-planar, has ten lines with two additional irreducible scalar products
and exhibits a four-particle cut only.
Integrals belonging to this topology are reduced to eleven master integrals,
which we choose to be also of the form of eq.~\eqref{HPLmasterForm} and
with the last two indices for the irreducible scalar products set to be zero,
namely,
\begin{align}\label{SeaSnakeBasis}
 S_{1} &= \epsilon^2 \text{TTS}_4(2,0,0,0,2,0,2,0,0,1,0,0), \nonumber\\
 S_{2} &= \epsilon^2 (1-x) \left[\text{TTS}_4(1,0,0,0,2,0,2,0,0,2,0,0)
                              +3 \text{TTS}_4(2,0,0,0,2,0,2,0,0,1,0,0)\right], \nonumber\\
 S_{3} &= \epsilon^3 \text{TTS}_4(2,0,1,0,1,0,2,0,0,1,0,0), \nonumber\\
 S_{4} &= \epsilon^4 \text{TTS}_4(2,0,0,1,1,1,1,0,0,1,0,0), \nonumber\\
 S_{5} &= \epsilon^4 \text{TTS}_4(1,2,1,0,1,0,1,0,0,1,0,0), \nonumber\\
 S_{6} &= \epsilon^4 \text{TTS}_4(2,1,1,0,1,0,1,0,0,1,0,0), \nonumber\\
 S_{7} &= \epsilon^3 (1-x) \text{TTS}_4(1,1,1,0,2,0,2,0,0,1,0,0), \nonumber\\
 S_{8} &= \epsilon^3 x \text{TTS}_4(1,2,2,0,1,0,1,0,0,1,0,0), \nonumber\\
 S_{9} &= \epsilon^3 x \text{TTS}_4(2,1,1,0,1,0,1,0,0,2,0,0), \nonumber\\
 S_{10} &= \epsilon^5 \text{TTS}_4(1,0,0,1,1,1,1,1,1,1,0,0), \nonumber\\
 S_{11} &= \epsilon^5 \text{TTS}_4(1,1,1,0,1,0,1,1,1,1,0,0).
\end{align}
Diagrams appearing in this basis are shown in fig.~\ref{fig:NNNLOdiags}.

The choice of master integrals $S_1$ and $S_2$ is motivated by the discussion given in
section~\ref{sec:NNLO} in analogy to eq.~\eqref{V1V2}.
$S_3$ and $S_4$ are not coupled and can be found by trying possible candidates.
$S_{10}$ and $S_{11}$ stem up to $\epsilon$-normalization from the reduction basis.
This can be motivated from
the sum of indices $a=8$ leading to the favored $e_W=1$ given
in eq.~\eqref{alphaRep}, as discussed in section~\ref{sec:NLO}.
Integrals $S_5$ to $S_9$ are coupled.
It is worth mentioning that the subtopology spanned by their six propagators
defines the $K_4$ topology discussed in ref.~\cite{Henn:2013nsa}.
Although kinematics and mass-configuration are different from the present case,
the basis integrals $S_5$ to $S_9$ can be established by direct diagrammatic
correspondence to five of the seven coupled integrals
for the off-shell $K_4$ case $g_6$, $\dots$, $g_{10}$ given in ref.~\cite{Henn:2013nsa}:
\begin{align}
  \label{correspondgiHj}
  g_6 \mathrel{\hat=} S_6, \qquad
  g_7 \mathrel{\hat=} S_5, \qquad
  g_8 \mathrel{\hat=} S_7, \qquad
  g_9 \mathrel{\hat=} S_9, \qquad
  g_{10} \mathrel{\hat=} S_8,
\end{align}
where ``$\mathrel{\hat=}$'' states that the indices of the integrals have the same
structure.
For this case of five coupled master integrals,
we do not use the algorithm described in section~\ref{subsubsec:3OrMore},
because the equations to be solved become highly complicated.

The basis given here satisfies the differential equation of eq.~\eqref{HennDEQ}
with eq.~\eqref{AForm}:
\begin{align}
a_4=
\begin{pmatrix}
 -4 & 1 & 0 & 0 & 0 & 0 & 0 & 0 & 0 & 0 & 0 \\
 -12 & 3 & 0 & 0 & 0 & 0 & 0 & 0 & 0 & 0 & 0 \\
 -\frac{1}{2} & 0 & -2 & 0 & 0 & 0 & 0 & 0 & 0 & 0 & 0 \\
 1 & -\frac{1}{6} & 0 & -3 & 0 & 0 & 0 & 0 & 0 & 0 & 0 \\
 -\frac{1}{6} & 0 & 0 & 0 & -\frac{2}{3} & \frac{2}{3} & -\frac{1}{6} & \frac{1}{2} & \frac{1}{3} & 0 & 0 \\
 \frac{5}{6} & 0 & 1 & 0 & \frac{7}{3} & -\frac{7}{3} & \frac{7}{12} & \frac{1}{4} & -\frac{1}{6} & 0 & 0 \\
 \frac{11}{3} & -\frac{1}{2} & 2 & 0 & \frac{14}{3} & -\frac{14}{3} & \frac{7}{6} & \frac{5}{2} & \frac{8}{3} & 0 & 0 \\
 \frac{7}{3} & -\frac{1}{2} & 2 & 0 & -\frac{2}{3} & \frac{2}{3} & -\frac{1}{6} & -\frac{7}{2} & -\frac{2}{3} & 0 & 0 \\
 \frac{7}{3} & -\frac{1}{2} & 2 & 0 & -\frac{2}{3} & \frac{2}{3} & -\frac{1}{6} & -\frac{1}{2} & -\frac{11}{3} & 0 & 0 \\
 0 & -\frac{1}{6} & 0 & 2 & 0 & 0 & 0 & 0 & 0 & -3 & 0 \\
 -3 & 0 & -7 & 0 & -13 & -2 & -\frac{3}{2} & -\frac{5}{2} & -2 & 0 & -3 \\
\end{pmatrix},
\end{align}
\begin{align}
b_4=
\begin{pmatrix}
 0 & 1 & 0 & 0 & 0 & 0 & 0 & 0 & 0 & 0 & 0 \\
 0 & 6 & 0 & 0 & 0 & 0 & 0 & 0 & 0 & 0 & 0 \\
 0 & 0 & 0 & 0 & 0 & 0 & 0 & 0 & 0 & 0 & 0 \\
 0 & 0 & 0 & 0 & 0 & 0 & 0 & 0 & 0 & 0 & 0 \\
 0 & 0 & 0 & 0 & 0 & 0 & 0 & 0 & 0 & 0 & 0 \\
 0 & 0 & 0 & 0 & 0 & 0 & 0 & 0 & 0 & 0 & 0 \\
 \frac{11}{3} & -\frac{1}{2} & 6 & 0 & \frac{50}{3} & \frac{10}{3} & \frac{37}{6} & \frac{7}{2} & \frac{5}{3} & 0 & 0 \\
 \frac{11}{3} & -\frac{1}{2} & 6 & 0 & \frac{50}{3} & \frac{10}{3} & \frac{1}{6} & \frac{7}{2} & \frac{5}{3} & 0 & 0 \\
 -\frac{11}{3} & -\frac{1}{2} & -6 & 0 & -\frac{50}{3} & -\frac{10}{3} & -\frac{19}{6} & -\frac{7}{2} & -\frac{5}{3} & 0 & 0 \\
 0 & 0 & 0 & 0 & 0 & 0 & 0 & 0 & 0 & 0 & 0 \\
 0 & 0 & 0 & 0 & 0 & 0 & 0 & 0 & 0 & 0 & 0 \\
\end{pmatrix}
\end{align}
and
\begin{align}
c_4=0.
\end{align}

We computed the master integrals
$S_i$ up to order $\epsilon^6$ corresponding to a maximum weight of six
in the HPLs.
As boundary conditions we computed the master integrals in the reduction basis
in the $x\to 1$ limit with our \prog{Mathematica} implementation of
the soft expansion algorithm given in ref.~\cite{Anastasiou:2013srw},
using the program \prog{FIRE} for reduction.
It is worth mentioning that all master integrals are reduced to
the four-particle phase space integral only, i.e.,
the integral $F_1(\epsilon)$ given in ref.~\cite{Anastasiou:2013srw}
which is known for general values of $\epsilon$, therefore
the result given here can be extended to arbitrary orders in $\epsilon$.
We checked our results transformed to the reduction basis against the soft expansion
around $x = 1$, including at least three non-vanishing terms
in the expansion.
Furthermore, we found that all $S_i$ are indeed
pure functions.
The results, up to order $\epsilon^5$ for brevity, are given in
appendix~\ref{app:HiResults}.

\section{Conclusions and outlook}
\label{sec:conclusion}

In this work we recomputed all necessary double-real and real-virtual master integrals
for $gg\to h$ at NNLO within the framework of canonical differential
equations~\cite{Henn:2013pwa}.
By solving a non-planar three-loop topology as well, we have explicitly shown
the use of the method to Higgs production in gluon fusion
in full $x$-dependence at \NNNLO{}.
The method can be applied to other master integrals at
\NNNLO{}, which will contribute to the completion of the calculation of
the total inclusive Higgs production cross section with full
$x$-dependence in the future.
To accomplish this, one needs to provide boundary conditions for
the differential equations.
For the triple-real emission diagrams,
the reduction of phase space integrals in the soft limit~%
\cite{Anastasiou:2013srw} works efficiently to obtain the soft expansions around
threshold.
For other diagrams containing loop integrals,
techniques such as asymptotic expansions by strategy of regions~%
\cite{Beneke:1997zp,Smirnov:2002pj,Anastasiou:2013mca},
those in the $\alpha$-parameter representations~%
\cite{Smirnov:1999bza,Smirnov:2012gma} or
asymptotic expansions of
Mellin-Barnes integral representations~\cite{Smirnov:2012gma}
may be useful to obtain the soft expansions.
Sophisticated treatments for phase space integrals may also do the job.
In any case, the fact that the framework of canonical differential equations
requires only the leading terms of expansions for boundary conditions
can make the computation simple.

Furthermore, working in this new framework
we were able to derive a new criterion to find a canonical master integral
being part of a coupled system
by use of a characteristic form for higher order differential equations,
allowing also for the construction of a canonical basis for the coupled integrals.
This criterion was derived in a quite general way and can also be used in
coupled systems of differential equations for master integrals
appearing in other processes.

\acknowledgments

We are grateful to J.M.~Henn for a comprehensive introduction to his method
as well as for pointing out the applicability of it to the process of
Higgs production in gluon fusion at \NNNLO{}.
We thank J.M.~Henn and
M.~Steinhauser for reading the manuscript and giving
useful comments.
Furthermore, we thank M.~Steinhauser, J.~Grigo and C.~Anzai
for valuable discussion.
We also thank A.V.~Smirnov for allowing us to use
the unpublished C++ version of \prog{FIRE}.
The Feynman diagrams were drawn by using \prog{JaxoDraw}~\cite{Binosi:2008ig},
based on \prog{Axodraw}~\cite{Vermaseren:1994je}.
This work was supported by Deutsche Forschungsgemeinschaft
through Sonderforschungsbereich/Transregio 9 "Computational Particle Physics".

\appendix

\section{Results for sea snake topology}
\label{app:HiResults}

In the appendix, we present the results of the master integrals $S_i$
for the sea snake topology $\text{TTS}_4$.
We have normalized the results by multiplying an $\epsilon$-dependent prefactor,
which is chosen such that the four-particle phase space integral in the limit
$x\to 1$, corresponding to $F_1(\epsilon)$ defined in
ref.~\cite{Anastasiou:2013srw}, becomes
\begin{align}
\lim_{x\to 1}\text{TTS}_4(1,0,0,0,1,0,1,0,0,1,0,0)=
  \frac{\Gamma^3 (1-\epsilon)}{\Gamma (6-6 \epsilon) \Gamma^3 (1+\epsilon)}.
\end{align}
Furthermore, we omit the common argument $x$ of HPLs as
$\H_{i_1,\dots i_n}:=\H_{i_1,\dots i_n}(x)$.
\begingroup
\allowdisplaybreaks[1]
\begin{align}
S_{1} &= 1 + 2 \epsilon \big[\H_{0} + 3 \H_{1}\big] +
       2 \epsilon^2 \big[6 \H_{2} - \H_{0, 0} + 3 \H_{1, 0} +
         18 \H_{1, 1} - 12 \zeta_2\big] - 2 \epsilon^3 \big[6 \H_{3} -
         18 \H_{1, 2}\nonumber\\
& - 6 \H_{2, 0} - 36 \H_{2, 1} -
         \H_{0, 0, 0} + 3 \H_{1, 0, 0} - 18 \H_{1, 1, 0} -
         108 \H_{1, 1, 1} + \big(6 \H_{0} + 54 \H_{1}\big) \zeta_2 +
         32 \zeta_3\big]\nonumber\\
& + 2 \epsilon^4 \big[6 \H_{4} - 18 \H_{1, 3} +
         36 \H_{2, 2} - 6 \H_{3, 0} - 36 \H_{3, 1} +
         108 \H_{1, 1, 2} + 18 \H_{1, 2, 0} +
         108 \H_{1, 2, 1}\nonumber\\
& - 6 \H_{2, 0, 0} + 36 \H_{2, 1, 0} +
         216 \H_{2, 1, 1} - \H_{0, 0, 0, 0} +
         3 \H_{1, 0, 0, 0} - 18 \H_{1, 1, 0, 0} +
         108 \H_{1, 1, 1, 0}\nonumber\\
& + 648 \H_{1, 1, 1, 1} +
         \big(-108 \H_{2} + 6 \H_{0, 0} - 18 \H_{1, 0} -
           324 \H_{1, 1}\big) \zeta_2 + 123 \zeta_4 +
         \big(-10 \H_{0}\nonumber\\
& - 138 \H_{1}\big) \zeta_3\big] -
       2 \epsilon^5 \big[6 \H_{5} - 18 \H_{1, 4} + 36 \H_{2, 3} +
         36 \H_{3, 2} - 6 \H_{4, 0} - 36 \H_{4, 1} +
         108 \H_{1, 1, 3}\nonumber\\
& - 108 \H_{1, 2, 2} +
         18 \H_{1, 3, 0} + 108 \H_{1, 3, 1} -
         216 \H_{2, 1, 2} - 36 \H_{2, 2, 0} -
         216 \H_{2, 2, 1} - 6 \H_{3, 0, 0}\nonumber\\
& + 36 \H_{3, 1, 0} +
         216 \H_{3, 1, 1} - 648 \H_{1, 1, 1, 2} -
         108 \H_{1, 1, 2, 0} - 648 \H_{1, 1, 2, 1} +
         18 \H_{1, 2, 0, 0}\nonumber\\
& - 108 \H_{1, 2, 1, 0} -
         648 \H_{1, 2, 1, 1} - 6 \H_{2, 0, 0, 0} +
         36 \H_{2, 1, 0, 0} - 216 \H_{2, 1, 1, 0} -
         1296 \H_{2, 1, 1, 1}\nonumber\\
& - \H_{0, 0, 0, 0, 0} +
         3 \H_{1, 0, 0, 0, 0} - 18 \H_{1, 1, 0, 0, 0} +
         108 \H_{1, 1, 1, 0, 0} - 648 \H_{1, 1, 1, 1, 0} -
         3888 \H_{1, 1, 1, 1, 1}\nonumber\\
& + \big(-21 \H_{0} - 513 \H_{1}\big)
          \zeta_4 + \zeta_2 \big(-108 \H_{3} + 324 \H_{1, 2} +
           36 \H_{2, 0} + 648 \H_{2, 1} + 6 \H_{0, 0, 0}\nonumber\\
& -
           18 \H_{1, 0, 0} + 108 \H_{1, 1, 0} +
           1944 \H_{1, 1, 1} - 444 \zeta_3\big) +
         \big(276 \H_{2} - 10 \H_{0, 0} + 30 \H_{1, 0}\nonumber\\
& +
           828 \H_{1, 1}\big) \zeta_3 + 324 \zeta_5\big] + \mathcal{O}\big(\epsilon^6\big),\\
     S_{2} &= 6 + 6 \epsilon \big[\H_{0} + 6 \H_{1}\big] +
       6 \epsilon^2 \big[6 \H_{2} - \H_{0, 0} + 6 \H_{1, 0} +
         36 \H_{1, 1} - 18 \zeta_2\big] - 6 \epsilon^3 \big[6 \H_{3} -
         36 \H_{1, 2}\nonumber\\
& - 6 \H_{2, 0} - 36 \H_{2, 1} -
         \H_{0, 0, 0} + 6 \H_{1, 0, 0} - 36 \H_{1, 1, 0} -
         216 \H_{1, 1, 1} + \big(6 \H_{0} + 108 \H_{1}\big) \zeta_2\nonumber\\
& +
         46 \zeta_3\big] + 6 \epsilon^4 \big[6 \H_{4} - 36 \H_{1, 3} +
         36 \H_{2, 2} - 6 \H_{3, 0} - 36 \H_{3, 1} +
         216 \H_{1, 1, 2} + 36 \H_{1, 2, 0}\nonumber\\
& +
         216 \H_{1, 2, 1} - 6 \H_{2, 0, 0} + 36 \H_{2, 1, 0} +
         216 \H_{2, 1, 1} - \H_{0, 0, 0, 0} +
         6 \H_{1, 0, 0, 0} - 36 \H_{1, 1, 0, 0}\nonumber\\
& +
         216 \H_{1, 1, 1, 0} + 1296 \H_{1, 1, 1, 1} +
         \big(-108 \H_{2} + 6 \H_{0, 0} - 36 \H_{1, 0} -
           648 \H_{1, 1}\big) \zeta_2 + 171 \zeta_4\nonumber\\
& +
         \big(-10 \H_{0} - 276 \H_{1}\big) \zeta_3\big] -
       6 \epsilon^5 \big[6 \H_{5} - 36 \H_{1, 4} + 36 \H_{2, 3} +
         36 \H_{3, 2} - 6 \H_{4, 0} - 36 \H_{4, 1}\nonumber\\
& +
         216 \H_{1, 1, 3} - 216 \H_{1, 2, 2} +
         36 \H_{1, 3, 0} + 216 \H_{1, 3, 1} -
         216 \H_{2, 1, 2} - 36 \H_{2, 2, 0} -
         216 \H_{2, 2, 1}\nonumber\\
& - 6 \H_{3, 0, 0} + 36 \H_{3, 1, 0} +
         216 \H_{3, 1, 1} - 1296 \H_{1, 1, 1, 2} -
         216 \H_{1, 1, 2, 0} - 1296 \H_{1, 1, 2, 1}\nonumber\\
& +
         36 \H_{1, 2, 0, 0} - 216 \H_{1, 2, 1, 0} -
         1296 \H_{1, 2, 1, 1} - 6 \H_{2, 0, 0, 0} +
         36 \H_{2, 1, 0, 0} - 216 \H_{2, 1, 1, 0}\nonumber\\
& -
         1296 \H_{2, 1, 1, 1} - \H_{0, 0, 0, 0, 0} +
         6 \H_{1, 0, 0, 0, 0} - 36 \H_{1, 1, 0, 0, 0} +
         216 \H_{1, 1, 1, 0, 0} - 1296 \H_{1, 1, 1, 1, 0}\nonumber\\
& -
         7776 \H_{1, 1, 1, 1, 1} + \big(-21 \H_{0} - 1026 \H_{1}\big)
          \zeta_4 + \zeta_2 \big(-108 \H_{3} + 648 \H_{1, 2} +
           36 \H_{2, 0} + 648 \H_{2, 1}\nonumber\\
& + 6 \H_{0, 0, 0} -
           36 \H_{1, 0, 0} + 216 \H_{1, 1, 0} +
           3888 \H_{1, 1, 1} - 612 \zeta_3\big) +
         \big(276 \H_{2} - 10 \H_{0, 0} + 60 \H_{1, 0}\nonumber\\
& +
           1656 \H_{1, 1}\big) \zeta_3 + 450 \zeta_5\big] + \mathcal{O}\big(\epsilon^6\big),\\
     S_{3} &= -\frac{\epsilon}{2} \H_{0} - 3 \epsilon^2 \big[\H_{2} - \zeta_2\big] +
       \epsilon^3 \big[-3 \H_{2, 0} - 18 \H_{2, 1} + \H_{0, 0, 0} +
         6 \H_{0} \zeta_2 + 12 \zeta_3\big]\nonumber\\
& +
       \epsilon^4 \big[6 \H_{4} - 18 \H_{2, 2} + 3 \H_{2, 0, 0} -
         18 \H_{2, 1, 0} - 108 \H_{2, 1, 1} -
         3 \H_{0, 0, 0, 0} + \big(54 \H_{2} - 6 \H_{0, 0}\big)
          \zeta_2\nonumber\\
& - 51 \zeta_4 + 8 \H_{0} \zeta_3\big] +
       \epsilon^5 \big[-18 \H_{5} + 18 \H_{2, 3} + 6 \H_{4, 0} +
         36 \H_{4, 1} - 108 \H_{2, 1, 2} - 18 \H_{2, 2, 0}\nonumber\\
& -
         108 \H_{2, 2, 1} - 3 \H_{2, 0, 0, 0} +
         18 \H_{2, 1, 0, 0} - 108 \H_{2, 1, 1, 0} -
         648 \H_{2, 1, 1, 1} + 7 \H_{0, 0, 0, 0, 0} -
         21 \H_{0} \zeta_4\nonumber\\
& + \zeta_2 \big(18 \H_{2, 0} +
           324 \H_{2, 1} + 6 \H_{0, 0, 0} - 192 \zeta_3\big) +
         \big(138 \H_{2} - 6 \H_{0, 0}\big) \zeta_3 + 138 \zeta_5\big] +
       \mathcal{O}\big(\epsilon^6\big), \\
S_{4} &= \epsilon^2 \H_{0, 0} +
       2 \epsilon^3 \big[3 \H_{3}
 - 2 \H_{0, 0, 0} - 3 \H_{0} \zeta_2 -
         3 \zeta_3\big] +
       \epsilon^4 \big[-24 \H_{4} + 6 \H_{3, 0} + 36 \H_{3, 1}\nonumber\\
& +
         13 \H_{0, 0, 0, 0} + 12 \H_{0, 0} \zeta_2 + 33 \zeta_4\big] + 2 \epsilon^5 \big[39 \H_{5} + 18 \H_{3, 2} -
         12 \H_{4, 0} - 72 \H_{4, 1} - 3 \H_{3, 0, 0}\nonumber\\
& +
         18 \H_{3, 1, 0} + 108 \H_{3, 1, 1} -
         20 \H_{0, 0, 0, 0, 0} - 12 \H_{0} \zeta_4 -
         5 \H_{0, 0} \zeta_3 + \zeta_2 \big(-54 \H_{3} -
           15 \H_{0, 0, 0}\nonumber\\
& + 36 \zeta_3\big) - 33 \zeta_5\big] +
       \mathcal{O}\big(\epsilon^6\big),\\
S_{5} &=
      \frac{\epsilon^3}{2} \big[\H_{2, 0} + \H_{0, 0, 0} + \H_{0} \zeta_2 +
          2 \zeta_3\big] + \epsilon^4 \big[3 \H_{4} + 3 \H_{2, 2} -
         2 \H_{3, 0} - \H_{2, 0, 0} + \H_{2, 1, 0}\nonumber\\
& -
         4 \H_{0, 0, 0, 0} + \big(-2 \H_{2} - 5 \H_{0, 0}\big)
          \zeta_2 - 2 \zeta_4 - \H_{0} \zeta_3\big] +
       \frac{\epsilon^5}{8} \big[-192 \H_{5} - 48 \H_{2, 3} - 96 \H_{3, 2}\nonumber\\
& +
          84 \H_{4, 0} + 144 \H_{4, 1} + 48 \H_{2, 1, 2} +
          4 \H_{2, 2, 0} + 144 \H_{2, 2, 1} +
          32 \H_{3, 0, 0} - 32 \H_{3, 1, 0}\nonumber\\
& -
          20 \H_{2, 0, 0, 0} - 48 \H_{2, 1, 0, 0} +
          172 \H_{0, 0, 0, 0, 0} + 323 \H_{0} \zeta_4 +
          \zeta_2 \big(64 \H_{3} - 20 \H_{2, 0} - 48 \H_{2, 1}\nonumber\\
& +
            204 \H_{0, 0, 0} - 56 \zeta_3\big) +
          \big(56 \H_{2} + 24 \H_{0, 0}\big) \zeta_3 + 184 \zeta_5\big] +
       \mathcal{O}\big(\epsilon^6\big),\\
S_{6} &=
      \frac{\epsilon^4}{4} \big[4 \H_{3, 0} + 8 \H_{2, 0, 0} + 4 \H_{2, 1, 0} +
          8 \H_{0, 0, 0, 0} + \big(4 \H_{2} + 4 \H_{0, 0}\big)
           \zeta_2 - 17 \zeta_4 - 4 \H_{0} \zeta_3\big]\nonumber\\
& +
       \frac{\epsilon^5}{2} \big[24 \H_{5} + 24 \H_{2, 3} + 12 \H_{3, 2} -
          12 \H_{4, 0} + 12 \H_{2, 1, 2} + 6 \H_{2, 2, 0} -
          16 \H_{3, 0, 0} - 2 \H_{3, 1, 0}\nonumber\\
& +
          20 \H_{2, 1, 0, 0} + 16 \H_{2, 1, 1, 0} -
          36 \H_{0, 0, 0, 0, 0} - 29 \H_{0} \zeta_4 +
          \big(-24 \H_{2} - 6 \H_{0, 0}\big) \zeta_3 +
          \zeta_2 \big(-14 \H_{3}\nonumber\\
& - 18 \H_{2, 0} + 4 \H_{2, 1} -
            36 \H_{0, 0, 0} + 10 \zeta_3\big) - 27 \zeta_5\big] +
       \mathcal{O}\big(\epsilon^6\big),\\
S_{7} &=
      -\frac{3}{2} + \frac{\epsilon}{4} \big[-7 \H_{0} - 36 \H_{1}\big] +
       \frac{\epsilon^2}{2} \big[-21 \H_{2} + 4 \H_{0, 0} - 19 \H_{1, 0} -
          108 \H_{1, 1} + 56 \zeta_2\big]\nonumber\\
& +
       \frac{\epsilon^3}{4} \big[48 \H_{3} - 228 \H_{1, 2} - 37 \H_{2, 0} -
          252 \H_{2, 1} - \H_{0, 0, 0} + 48 \H_{1, 0, 0} -
          216 \H_{1, 1, 0} - 1296 \H_{1, 1, 1}\nonumber\\
& +
          \big(41 \H_{0} + 660 \H_{1}\big) \zeta_2 + 262 \zeta_3\big] +
       \frac{\epsilon^4}{8} \big[-12 \H_{4} + 576 \H_{1, 3} - 444 \H_{2, 2} +
          72 \H_{3, 0}\nonumber\\
& + 576 \H_{3, 1} - 2592 \H_{1, 1, 2} -
          412 \H_{1, 2, 0} - 2736 \H_{1, 2, 1} +
          80 \H_{2, 0, 0} - 476 \H_{2, 1, 0} -
          3024 \H_{2, 1, 1}\nonumber\\
& - 68 \H_{0, 0, 0, 0} -
          52 \H_{1, 0, 0, 0} + 560 \H_{1, 1, 0, 0} -
          2528 \H_{1, 1, 1, 0} - 15552 \H_{1, 1, 1, 1} +
          \big(1480 \H_{2}\nonumber\\
& - 204 \H_{0, 0} + 380 \H_{1, 0} +
            7840 \H_{1, 1}\big) \zeta_2 - 2547 \zeta_4 +
          \big(68 \H_{0} + 3160 \H_{1}\big) \zeta_3\big] +
       \frac{\epsilon^5}{16} \big[-816 \H_{5}\nonumber\\
& - 624 \H_{1, 4} + 960 \H_{2, 3} +
          864 \H_{3, 2} + 84 \H_{4, 0} - 144 \H_{4, 1} +
          6720 \H_{1, 1, 3} - 4944 \H_{1, 2, 2}\nonumber\\
& +
          944 \H_{1, 3, 0} + 6912 \H_{1, 3, 1} -
          5712 \H_{2, 1, 2} - 940 \H_{2, 2, 0} -
          5328 \H_{2, 2, 1} - 192 \H_{3, 0, 0}\nonumber\\
& +
          1008 \H_{3, 1, 0} + 6912 \H_{3, 1, 1} -
          30336 \H_{1, 1, 1, 2} - 4800 \H_{1, 1, 2, 0} -
          31104 \H_{1, 1, 2, 1} + 864 \H_{1, 2, 0, 0}\nonumber\\
& -
          5232 \H_{1, 2, 1, 0} - 32832 \H_{1, 2, 1, 1} -
          268 \H_{2, 0, 0, 0} + 928 \H_{2, 1, 0, 0} -
          5920 \H_{2, 1, 1, 0} - 36288 \H_{2, 1, 1, 1}\nonumber\\
& +
          596 \H_{0, 0, 0, 0, 0} - 464 \H_{1, 0, 0, 0, 0} -
          864 \H_{1, 1, 0, 0, 0} + 6464 \H_{1, 1, 1, 0, 0} -
          30080 \H_{1, 1, 1, 1, 0}\nonumber\\
& - 186624 \H_{1, 1, 1, 1, 1} +
          \big(1343 \H_{0} - 27600 \H_{1}\big) \zeta_4 +
          \zeta_2 \big(-3312 \H_{3} + 16128 \H_{1, 2} +
            764 \H_{2, 0}\nonumber\\
& + 17936 \H_{2, 1} +
            972 \H_{0, 0, 0} - 1888 \H_{1, 0, 0} +
            4032 \H_{1, 1, 0} + 93568 \H_{1, 1, 1} -
            13352 \zeta_3\big)\nonumber\\
& + \big(7816 \H_{2} - 200 \H_{0, 0} +
            720 \H_{1, 0} + 38208 \H_{1, 1}\big) \zeta_3 +
          8976 \zeta_5\big] + \mathcal{O}\big(\epsilon^6\big),\\
     S_{8} &= -\frac{\epsilon}{4} \H_{0} +
       \frac{\epsilon^2}{2} \big[-3 \H_{2} + 3 \H_{0, 0} + 2 \H_{1, 0} +
          5 \zeta_2\big] + \frac{\epsilon^3}{4} \big[36 \H_{3} + 24 \H_{1, 2} -
          19 \H_{2, 0}\nonumber\\
& - 36 \H_{2, 1} - 27 \H_{0, 0, 0} +
          12 \H_{1, 1, 0} + \big(-37 \H_{0} - 12 \H_{1}\big) \zeta_2 -
          2 \zeta_3\big] + \frac{\epsilon^4}{4} \big[-162 \H_{4}\nonumber\\
& - 114 \H_{2, 2} +
          78 \H_{3, 0} + 216 \H_{3, 1} + 72 \H_{1, 1, 2} +
          16 \H_{1, 2, 0} + 144 \H_{1, 2, 1} +
          8 \H_{2, 0, 0} - 74 \H_{2, 1, 0}\nonumber\\
& -
          216 \H_{2, 1, 1} + 108 \H_{0, 0, 0, 0} -
          20 \H_{1, 0, 0, 0} - 8 \H_{1, 1, 0, 0} +
          32 \H_{1, 1, 1, 0} + \big(148 \H_{2} + 132 \H_{0, 0}\nonumber\\
& -
            56 \H_{1, 0} - 40 \H_{1, 1}\big) \zeta_2 + 215 \zeta_4 +
          \big(10 \H_{0} + 8 \H_{1}\big) \zeta_3\big] +
       \frac{\epsilon^5}{16} \big[2592 \H_{5} - 480 \H_{1, 4}\nonumber\\
& + 192 \H_{2, 3} +
          1872 \H_{3, 2} - 1188 \H_{4, 0} - 3888 \H_{4, 1} -
          192 \H_{1, 1, 3} + 384 \H_{1, 2, 2} +
          80 \H_{1, 3, 0}\nonumber\\
& - 1776 \H_{2, 1, 2} -
          340 \H_{2, 2, 0} - 2736 \H_{2, 2, 1} -
          192 \H_{3, 0, 0} + 1344 \H_{3, 1, 0} +
          5184 \H_{3, 1, 1}\nonumber\\
& + 768 \H_{1, 1, 1, 2} +
          144 \H_{1, 1, 2, 0} + 1728 \H_{1, 1, 2, 1} -
          96 \H_{1, 2, 0, 0} + 480 \H_{1, 2, 1, 0} +
          3456 \H_{1, 2, 1, 1}\nonumber\\
& + 236 \H_{2, 0, 0, 0} +
          288 \H_{2, 1, 0, 0} - 1248 \H_{2, 1, 1, 0} -
          5184 \H_{2, 1, 1, 1} - 1620 \H_{0, 0, 0, 0, 0} +
          352 \H_{1, 0, 0, 0, 0}\nonumber\\
& - 240 \H_{1, 1, 0, 0, 0} -
          256 \H_{1, 1, 1, 0, 0} + 256 \H_{1, 1, 1, 1, 0} +
          \big(-2935 \H_{0} + 2964 \H_{1}\big) \zeta_4 +
          \zeta_2 \big(-3120 \H_{3}\nonumber\\
& - 1632 \H_{1, 2} +
            836 \H_{2, 0} + 3120 \H_{2, 1} -
            1836 \H_{0, 0, 0} + 560 \H_{1, 0, 0} -
            528 \H_{1, 1, 0} - 512 \H_{1, 1, 1}\nonumber\\
& - 264 \zeta_3\big) +
          \big(952 \H_{2} - 168 \H_{0, 0} - 96 \H_{1, 0} +
            288 \H_{1, 1}\big) \zeta_3 - 176 \zeta_5\big] +
       \mathcal{O}\big(\epsilon^6\big),\\
S_{9} &=
      -\frac{1}{4} + \frac{\epsilon}{2} \big[\H_{0} - 3 \H_{1}\big] +
       \frac{\epsilon^2}{4} \big[12 \H_{2} - 3 \H_{0, 0} - 7 \H_{1, 0} -
          36 \H_{1, 1} - \zeta_2\big] +
       \frac{\epsilon^3}{2} \big[-9 \H_{3}\nonumber\\
& - 21 \H_{1, 2} + 7 \H_{2, 0} +
          36 \H_{2, 1} - 21 \H_{1, 1, 0} - 108 \H_{1, 1, 1} +
          \big(-2 \H_{0} + 54 \H_{1}\big) \zeta_2 + \zeta_3\big]\nonumber\\
& +
       \frac{\epsilon^4}{16} \big[336 \H_{2, 2} - 84 \H_{3, 0} - 432 \H_{3, 1} -
          1008 \H_{1, 1, 2} - 196 \H_{1, 2, 0} -
          1008 \H_{1, 2, 1} + 32 \H_{2, 0, 0}\nonumber\\
& +
          352 \H_{2, 1, 0} + 1728 \H_{2, 1, 1} +
          108 \H_{0, 0, 0, 0} - 4 \H_{1, 0, 0, 0} +
          32 \H_{1, 1, 0, 0} - 992 \H_{1, 1, 1, 0}\nonumber\\
& -
          5184 \H_{1, 1, 1, 1} + \big(-848 \H_{2} +
            132 \H_{0, 0} + 308 \H_{1, 0} + 2608 \H_{1, 1}\big)
           \zeta_2 + 141 \zeta_4 + \big(-8 \H_{0}\nonumber\\
& + 1240 \H_{1}\big)
           \zeta_3\big] + \frac{\epsilon^5}{8} \big[324 \H_{5} - 12 \H_{1, 4} +
          96 \H_{2, 3} - 252 \H_{3, 2} + 96 \H_{1, 1, 3} -
          588 \H_{1, 2, 2}\nonumber\\
& + 32 \H_{1, 3, 0} +
          1056 \H_{2, 1, 2} + 220 \H_{2, 2, 0} +
          1008 \H_{2, 2, 1} - 96 \H_{3, 0, 0} -
          300 \H_{3, 1, 0} - 1296 \H_{3, 1, 1}\nonumber\\
& -
          2976 \H_{1, 1, 1, 2} - 564 \H_{1, 1, 2, 0} -
          3024 \H_{1, 1, 2, 1} + 96 \H_{1, 2, 0, 0} -
          540 \H_{1, 2, 1, 0} - 3024 \H_{1, 2, 1, 1}\nonumber\\
& +
          4 \H_{2, 0, 0, 0} + 48 \H_{2, 1, 0, 0} +
          1056 \H_{2, 1, 1, 0} + 5184 \H_{2, 1, 1, 1} -
          324 \H_{0, 0, 0, 0, 0} + 76 \H_{1, 0, 0, 0, 0}\nonumber\\
& -
          12 \H_{1, 1, 0, 0, 0} + 176 \H_{1, 1, 1, 0, 0} -
          2912 \H_{1, 1, 1, 1, 0} - 15552 \H_{1, 1, 1, 1, 1} +
          \big(-389 \H_{0}\nonumber\\
& - 2049 \H_{1}\big) \zeta_4 +
          \big(-1336 \H_{2} - 12 \H_{0, 0} + 324 \H_{1, 0} +
            3624 \H_{1, 1}\big) \zeta_3 + \zeta_2 \big(600 \H_{3}\nonumber\\
& +
            1560 \H_{1, 2} - 380 \H_{2, 0} - 2592 \H_{2, 1} -
            324 \H_{0, 0, 0} + 44 \H_{1, 0, 0} +
            852 \H_{1, 1, 0} + 7840 \H_{1, 1, 1}\nonumber\\
& + 4 \zeta_3\big) -
          88 \zeta_5\big] + \mathcal{O}\big(\epsilon^6\big),\\
     S_{10} &= -\epsilon \H_{0} - 2 \epsilon^2 \big[3 \H_{2} -
         \H_{0, 0} - 3 \zeta_2\big] + 3 \epsilon^3 \big[4 \H_{3} -
         2 \H_{2, 0} - 12 \H_{2, 1} - \H_{0, 0, 0} +
         4 \zeta_3\big]\nonumber\\
& - 2 \epsilon^4 \big[9 \H_{4} + 18 \H_{2, 2} -
         6 \H_{3, 0} - 36 \H_{3, 1} - 3 \H_{2, 0, 0} +
         18 \H_{2, 1, 0} + 108 \H_{2, 1, 1} +
         \big(-54 \H_{2}\nonumber\\
& + 3 \H_{0, 0}\big) \zeta_2 + 27 \zeta_4 +
         \H_{0} \zeta_3\big] + \epsilon^5 \big[36 \H_{2, 3} + 72 \H_{3, 2} -
         18 \H_{4, 0} - 108 \H_{4, 1} - 216 \H_{2, 1, 2}\nonumber\\
& -
         36 \H_{2, 2, 0} - 216 \H_{2, 2, 1} -
         12 \H_{3, 0, 0} + 72 \H_{3, 1, 0} +
         432 \H_{3, 1, 1} - 6 \H_{2, 0, 0, 0} +
         36 \H_{2, 1, 0, 0}\nonumber\\
& - 216 \H_{2, 1, 1, 0} -
         1296 \H_{2, 1, 1, 1} + 27 \H_{0, 0, 0, 0, 0} +
         57 \H_{0} \zeta_4 + \zeta_2 \big(-216 \H_{3} +
           36 \H_{2, 0}\nonumber\\
& + 648 \H_{2, 1} + 36 \H_{0, 0, 0} -
           132 \zeta_3\big) + \big(276 \H_{2} + 16 \H_{0, 0}\big) \zeta_3 +
         120 \zeta_5\big] + \mathcal{O}\big(\epsilon^6\big),\\
S_{11} &=
      -\frac{\epsilon}{4} \H_{0} + \frac{\epsilon^2}{2} \big[-3 \H_{2} + \H_{0, 0} +
          3 \zeta_2\big] + \frac{\epsilon^3}{4} \big[12 \H_{3} - 11 \H_{2, 0} -
          36 \H_{2, 1} - 3 \H_{0, 0, 0}\nonumber\\
& - 5 \H_{0} \zeta_2 +
          2 \zeta_3\big] + \frac{\epsilon^4}{2} \big[-9 \H_{4} - 33 \H_{2, 2} +
          11 \H_{3, 0} + 36 \H_{3, 1} - 27 \H_{2, 1, 0} -
          108 \H_{2, 1, 1}\nonumber\\
& + \big(60 \H_{2} + 2 \H_{0, 0}\big)
           \zeta_2 - 18 \zeta_4 - 9 \H_{0} \zeta_3\big] +
       \frac{\epsilon^5}{16} \big[528 \H_{3, 2} - 132 \H_{4, 0} - 432 \H_{4, 1}\nonumber\\
& -
          1296 \H_{2, 1, 2} - 260 \H_{2, 2, 0} -
          1584 \H_{2, 2, 1} + 432 \H_{3, 1, 0} +
          1728 \H_{3, 1, 1} + 76 \H_{2, 0, 0, 0}\nonumber\\
& +
          64 \H_{2, 1, 0, 0} - 1120 \H_{2, 1, 1, 0} -
          5184 \H_{2, 1, 1, 1} + 108 \H_{0, 0, 0, 0, 0} +
          97 \H_{0} \zeta_4 + \big(1208 \H_{2}\nonumber\\
& + 232 \H_{0, 0}\big)
           \zeta_3 + \zeta_2 \big(-960 \H_{3} + 532 \H_{2, 0} +
            2768 \H_{2, 1} + 84 \H_{0, 0, 0} + 456 \zeta_3\big) -
          200 \zeta_5\big]\nonumber\\
& + \mathcal{O}\big(\epsilon^6\big)
.
\end{align}
\endgroup

\bibliographystyle{JHEP}
\bibliography{gghbasis}

\end{document}